\documentclass {article}
\usepackage{graphicx}

\begin{document}

\title{On the origin of the matter-antimatter asymmetry in self-gravitating systems at ultra-high temperatures}

\author{Michael Petri\thanks{email: mpetri@bfs.de} \\Bundesamt f\"{u}r Strahlenschutz (BfS), Salzgitter, Germany}

\date{May 1, 2004}

\maketitle

\begin{abstract}
It is shown, that self-gravitating systems can be classified by a
dimensionless constant positive number $\kappa = S T / E$, which
can be determined from the (global) values for the entropy,
temperature and (total) energy. The Kerr-Newman black hole family
is characterized by $\kappa$ in the range $0-1/2$, depending on
the dimensionless ratios of angular momentum and charge squared to
the horizon area, $J/A$ and $Q^2/A$.

By analyzing the most general case of an ultra-relativistic ideal
gas with non-zero chemical potential it is shown, that $\kappa$ is
an important parameter which determines the (local) thermodynamic
properties of an ultra-relativistic gas. $\kappa$ only depends on
the chemical potential per temperature $u = \mu / T$ and on the
ratio of bosonic to fermionic degrees of freedom $r_F = f_B /
f_F$. A gas with zero chemical potential has $\kappa = 4/3$.
Whenever $\kappa < 4/3$ the gas must acquire a non-zero chemical
potential. This non-zero chemical potential induces a natural
matter-antimatter asymmetry, whenever microscopic statistical
thermodynamics can be applied.

The recently discovered holographic solution describes a compact
self gravitating black hole type object with an interior, well
defined matter state. One can associate a local - possibly
observer-dependent - value of $\kappa$ to the interior matter,
which lies in the range $2/3-1$ (for the uncharged case). This
finding is used to construct an alternative scenario of
baryogenesis in the context of the holographic solution, based on
quasi-equilibrium thermodynamics.

\end{abstract}

\maketitle

\section{\label{sec:intro}Introduction}

In our universe we experience a profound matter-antimatter
asymmetry. It's fundamental origin is not known. The standard
explanation for this asymmetry is attributed to the dynamic
evolution of the universe shortly after the "big-bang". According
to the mechanism first sketched out by Sacharov already in the
year 1967 \cite{Sakharov/1967}, a CP-violating process taking
place at high temperatures (such as the asymmetric decay of the
$X$ and $Y$ bosons below the GUT-scale) combined with a temporary
deviation from thermal equilibrium, as could have been caused by
the rapid expansion of the universe, can transform a slight
matter-antimatter asymmetry at high temperatures into a profound
asymmetry at low temperatures.

The low baryon to photon ratio of $\eta \approx 10^{-9}$
encountered in our universe in its present state is usually
interpreted as a remnant of a former minuscule asymmetry in the
baryon-antibaryon-number of the order of $10^{-9}$. According to
the common belief there were roughly $10^9 +1$ baryons vs. $10^9$
antibaryons at the time, when the temperature of the universe fell
below the rest-mass of the nucleon.\footnote{At a temperature of
$T \approx 1 \, GeV$ we expect a quark-gluon plasma. So it might
be more appropriate to talk of an asymmetry in the quark
anti-quark (and lepton/antilepton) pairs with mutual annihilation
of quarks and anti-quarks taking place at a somewhat lower
temperature. However, whether it is the nucleons that annihilate
at the nucleon threshold, or the quarks at roughly the
pion-threshold doesn't really change the basic picture.} The
baryons/antibaryons annihilated at this threshold, predominantly
into photons, leaving 1 baryon and $10^9$ photons behind. This
primordial ratio of $10^9$ photons per baryon then was preserved -
at least approximately - during the subsequent expansion.

Although such a scenario is thinkable, in the sense that it
doesn't appear to be in direct contradiction to any fundamental
physical laws, and is furthermore supported by (indirect)
experimental evidence such as today's high value for the photon to
baryon ratio which appears to fit well with the ratio predicted
from primordial nucleosynthesis\footnote{If one assumes that the
WMAP-determination of the baryonic matter content $\Omega_b
\approx 0.046$ is correct, the prediction of $\Omega_b$ from
primordial nucleosynthesis and the WMAP-value agree better than
50\%. However, there are some problems with respect to the
relative abundances of H to He4 to D to Li7. Whereas the He4/H and
Li7/H abundances indicate a common value of $\eta = n_b / n_\gamma
\approx 5 \cdot 10^{-10}$, the D/H-ratio requires a higher value,
which lies outside the error bars of the former value. See
\cite{Coc/2004} and the references therein for a detailed
analysis.}, the scenario relies on several implicit assumptions,
which appear questionable.

The first assumption is, that our universe is accurately described
by a homogeneous Friedman Robertson Walker (FRW) model, over the
full temperature range from the GUT-energy-breaking scale to the
low energy scale today. This assumption has been quite successful
in explaining many of the phenomena encountered in the observable
universe today in terms of a solution of the field equations of
only moderate mathematical complexity. On the other hand, today's
standard cosmological has lost much of its original charm. It has
turned into a complex model, relying on the experimental
determination of several crucial parameters (such as the fraction
of cold dark matter, dark energy) whose numerical values cannot be
predicted by today's methods and whose fundamental origin is not
known. There is {\em no theorem} that the universe {\em must} be
describable by an FRW-model.\footnote{If we believe in inflation,
the universe as a whole is chaotic and we happen to live in one of
its fairly homogeneous sub-compartments. According to string
theorists any sub-compartment with a positive value of the
cosmological constant has a very low probability of occurrence.}
In fact, the so called holographic solution, an exact solution to
the Einstein-field equations with zero cosmological constant, was
shown to be a potential alternative model for the universe
\cite{petri/hol} . In contrast to the FRW-model the holographic
solution has no free parameters. Yet it's unique properties fit
today's observational facts very well.\footnote{The holographic
solution is nearly indistinguishable from a homogeneously
expanding FRW-model at low energies and late times. In contrast to
the FRW-model, which has at least three free parameters $H,
\Omega_m, \Omega_V$, there is only one free "parameter" in the
holographic solution: the radial position $r$ of a geodesically
moving observer. $r$ has the meaning of a scale factor (or
alternatively: curvature radius) and is proportional to the proper
time $t$ measured by a geodesically moving observer, when
traveling from the hot central region (of Planck-density) to the
low density region today. From $t$ all other relevant cosmological
parameters, such as the local scale factor $r$, the current
Hubble-value $H = 1/t \simeq 1/r$, the local value of the
microwave-background temperature $T = \hbar / (4 \pi \sqrt{r_0
r})$ (with $r_0^2 \simeq 2 \sqrt{3} \hbar$), the total local
matter-density $\rho = 1 / (8 \pi r^2)$ etc. follow. These
parameters are related by non-trivial relations, such as $H t = 1$
and $T^4 / \rho = (\hbar / (4 \pi))^3 / \sqrt{3}$ (in units $c = G
= k = 1$). Remarkably, all relations predicted from the holostar
model are fulfilled to an accuracy of a few percent in the
observable universe today.} But its evolution very much differs
from that of a FRW-universe.

The second assumption is tied to the first. The standard scenario
of big-bang baryogenesis makes heavy use of the argument, that in
a homogeneously expanding FRW-type universe the particle numbers
of the different species in a co-moving volume-element should
remain constant during the expansion. This assumption is
model-dependent. Although it is true in the FRW-model, it can fail
in other models. In the holographic solution the baryon to photon
ratio is time dependent and evolves linearly with
temperature.\footnote{To be more precise: $n_b / n_\gamma \propto
T / m$ in the matter-dominated era. $m$ is the mass of a
fundamental particle, such as the electron or proton. This mass
must not necessarily be constant. It can be an arbitrary function
of temperature. Therefore the number ratio of baryons to photons
only depends linearly on the temperature, when the particle mass
is independent of temperature (or radius) in the interior holostar
space-time. Yet if the radiation temperature is referenced to the
mass of a fundamental particle, we have a linear dependence. The
reason for the different evolution of baryon- and photon-number
with temperature (or rather with the ratio $T/m$) in the
holographic universe is, that in the holographic solution is is
not relative particle numbers which is conserved during the
expansion, but rather the relative energy- and entropy-densities.
If one extrapolates this dependence back to $T \approx m_e$, one
gets the remarkable result that baryon to photon ratio in the
holographic solution is nearly unity at the electron-mass
threshold, i.e. when $T / m_e \approx $. This points to a
thermodynamic origin of this ratio.}

The third assumption is about the nature of the phase-transition
at the time of baryogenesis. The implicit assumption in the
standard cosmological model is, that the phase transition caused a
vast imbalance in particle numbers almost instantly: If one
compares the temperature-dependencies of the particle interaction
rates with the Hubble rate, one can conclude quite confidently
that the very early universe must have been in thermodynamic
equilibrium.\footnote{The reaction rates, which are proportional
to the number-densities of the interacting species, grow stronger
with $T$ than the Hubble rate in the radiation dominated phase.}
This means, that the number- and energy-densities of all known
particle species, such as baryons and photons, must have been
roughly equal at the time slightly before baryogenesis. Slightly
after baryogenesis the standard cosmological model, however,
postulates a discrepancy in the energy- and number-densities of
photons and (left over) nucleons of the order of $10^{9}$. This
extreme imbalance in particle numbers then is assumed to have been
preserved throughout the whole intermediate energy range up to
nucleosynthesis and beyond, down to the low temperatures
encountered today.

The justification for this type of phase transition is, that it
provides a plausible explanation for the profound matter
anti-matter asymmetry in our universe at low temperatures, if one
assumes a minuscule asymmetry at high temperatures (as the high
ratio of photons to baryons observed today seems to imply).
Unfortunately there is no mechanism which comes close to explain
the small primordial asymmetry of order $10^{-9}$. As long as this
crucial assumption, which lies at the heart of the matter
anti-matter problem, has not found a satisfactory theoretical
explanation, it is worthwhile to explore alternatives.

In this paper I attempt to give an alternative explanation for the
matter-antimatter asymmetry, which is based on equilibrium
thermodynamics of an ultra-relativistic gas of fermions and
bosons. The crucial observation is, that whenever the
ultra-relativistic fermions develop a {\em non-zero chemical
potential} comparable to the temperature, this can act as a
natural - purely thermodynamic - cause for a {\em profound}
matter-antimatter asymmetry at high temperatures.

That a non-zero chemical potential induces a matter-antimatter
asymmetry is a well known fact from microscopic statistical
thermodynamics. The question is, under what circumstances such a
non-zero chemical potential can arise. It turns out that
self-gravitating systems, which are characterized by the property
that their entropy can be expressed as a function of the energy
alone, i.e. $S = S(E)$, provide a natural setting for a non-zero
chemical potential of the fermions at ultra-relativistic
temperatures. Such systems are characterized by a strict
proportionality between the total energy and the free energy $F =
(1-\kappa) E$. The constant value of $\kappa$ can only take on a
very narrow range: $0 < \kappa \leq 4/3$. The standard case of an
ultra-relativistic gas with zero chemical potential is described
by $\kappa = 4/3$. Whenever $\kappa \neq 4/3$, non-zero values for
the chemical potentials of the fermions necessarily arise.
Classical black holes have values of $\kappa$ in the range $0 <
\kappa \leq 1/2$ ($\kappa = 1/2$ for a Schwarzschild black hole,
$\kappa \rightarrow 0$ for an extreme Kerr-Newman black hole). An
interesting value is $\kappa=1$, which is realized within the
holographic solution. For $\kappa=1$ the free energy is minimized
to zero.

The question, by what physical process the non-zero chemical
potentials of the fermions can arise in the first place, will not
be answered in this paper. It seems clear, that one requires some
(local) violation of $CP$ and/or $P$. In order to save the
$CPT$-theorem $T$ would have to be violated locally as well. The
weak interactions are known to violate $P$ maximally. $CP$ is
violated in certain weak decay processes. Quite interestingly, the
rotating holographic solution appears to provide a natural setting
for a significant local $CP$-violation of the macroscopic state,
if the interior ultra-relativistic particles (such as neutrinos or
anti-neutrinos) are aligned along the direction of the exterior
rotation axis.\footnote{This alignment also induces a substantial
local violation of $T$, as the primary direction of motion of
neutrinos and anti-neutrinos in the holographic solution differs
in the two half-spheres defined by the exterior axis. Any particle
in the holostar solution must acquire a highly relativistic,
nearly radial motion. If the neutrinos only have one helicity
state (as assumed in the Standard Model of particle physics) the
neutrinos (with spin opposite to their direction of flight) will
preferentially move outward in one half-sphere, whereas the
anti-neutrinos (with spin in direction of flight) will
preferentially move inward. For the other half-sphere the
situation is reversed.}  See \cite{petri/thermo} for more details.

The paper is divided into five principal sections. In section 2
the thermodynamic properties of the most general case of an
ultra-relativistic gas of fermions and bosons is discussed. A
relation between the chemical potential per temperature of the gas
and the dimensionless ratio $\kappa = S T / E$ of the system will
be derived. In section 3 I will discuss the relation between
energy and entropy for several self-gravitating systems and will
demonstrate, how the dimensionless ratio $\kappa$ determines the
global and local properties of the system. In section 4 necessary
conditions are discussed, under which an ultra-relativistic gas
can develop a non-zero chemical potential. In section 5 the
findings of the previous sections will be discussed for for some
particular self-gravitating systems. Section 6 then describes an
alternative scenario for the origin of the matter-antimatter
asymmetry, using the holostar solution as a simple model.

\section{Thermodynamics of an ultra-relativistic gas of fermions and bosons}

The objective of this section is to derive the thermodynamic
properties for the most general case of an ultra-relativistic
ideal gas consisting out of bosons and fermions. Most of the
explicit derivations will be done for fermions. The results then
are extended for the more general case of a gas consisting out of
bosons, fermions and anti-fermions.

\subsection{\label{sec:properties:gas}General properties of an ultra-relativistic ideal gas}

An ultra-relativistic gas must be described by the grand canonical
ensemble: At ultra-relativistic temperatures we will have copious
particle-interchange reactions between the different particle
species. Each species exchanges particles, energy and entropy with
the other species. The number of particles within any given
species cannot be considered fixed, but rather has to adjust to
the thermodynamic constraints.

Let us assume, that the time-scale of the interaction processes is
small enough so that thermal equilibrium can be attained, yet that
the interactions are weak enough so that the gas can be described - at
least approximately - as an ideal gas of massless, essentially
non-interacting particles. These assumptions should be valid at
the high temperatures and densities encountered in the very early
universe.

Under the ideal gas assumption the contribution of the individual
particle species to the extrinsic quantities, such as entropy and
energy, can be calculated separately. The total for each
(extrinsic) quantity is formed just by summing up the individual
contributions.

I will first consider a gas consisting only out of fermions. The
bosons, anti-bosons and anti-fermions will be added later. This
only changes some numerical factors, not the general picture. For
convenience units $c=1$ will be used throughout this paper.

A gas at ultra-high temperatures is expected to consist
exclusively out of elementary (i.e. not composed) particles. The
main characteristics, by which elementary particles can be
distinguished from each other are mass, spin and charge(s). At
ultra-high temperatures some of the characteristic features
distinguishing different particle species from each other will
become blurred: The particles will act more and more like truly
massless particles, the higher the temperature
becomes.\footnote{If it is the Higgs-mechanism that gives the
particles their masses, the particles will actually be truly
massless above the energy scale of the symmetry-breaking invoked
by the Higgs.} At ultra-high energies the particle's rest mass is
an utterly insignificant correction to the energy-momentum
relation $E \simeq p$. The actual value of the spin of a particle
doesn't play a significant role at ultra-high temperatures either,
as all transverse spin-directions are heavily
suppressed.\footnote{An ultra-relativistic particle effectively
has only two helicity-components, regardless of the number of
(transverse) spin-components in its rest-frame.} We might be able
to distinguish the particles by their different charges. However,
if the $GUT$-picture is correct, all known charges (electro-weak,
strong) will unify at some high energy, so that the charge looses
it's distinguishing quality at high energies. Furthermore the
fine-structure constant, as well as the other coupling constants,
are expected to remain small, even at the Planck energy, so that
the electric charge - as well as the other charges - only provide
a very moderate correction to the ideal gas law. If the gas is
neutral, the charge(s) of the particles most likely will be quite
irrelevant for the thermodynamic properties of a gas at ultra-high
temperatures.

Therefore, at energies well above the electro-weak scale it is not
unreasonable to assume that the particles are only distinguished
by the different representations of the Poincare-group for a
massless particle. The only label of a (non-tachyonic) particle in
the massless sector of the Poincare-group is the particle's
helicity. If the macroscopic state of the system does not single
out a preferred axis (no rotation), the spin of a particle is only
relevant to the thermodynamic description in the sense, that the
two opposite spin-components provide separate degrees of freedom
that must be included in the counting of the fundamental degrees
of freedom.\footnote{If there is a preferred axis, we have to
consider the spin-alignment of the particles with respect to this
axis. The microscopic energy usually depends on the product of the
particle's spin vector with the exterior axis, so that we have to
know the magnitude of spin-quantum number of the particles for a
complete thermodynamic description, whenever spherical symmetry is
broken.}

Therefore the only relevant {\em microscopic} parameter describing a
neutral, isotropic gas of ultra-relativistic fermions in
thermodynamic equilibrium should be the number of degrees of
freedom of the fermions. Let us denote this number by $f_F$.

However, we also have to consider the {\em macroscopic}
thermodynamic parameters. Therefore we cannot rule out the
possibility that the fermions have a non-zero chemical potential.
If this happens to be the case, it is reasonable to conjecture
that all fermionic species have the {\em same} chemical potential at
ultra-relativistic energies: The chemical potential is a measure
of how much energy must be invested to add a new particle to a
closed system, without changing its entropy or volume $\mu =
(\partial E / \partial N)_{S, V}$. At ultra-high temperatures,
where the rest-masses, charges etc. of the fermions are utterly
negligible, the energy required to add a new particle to the gas
is expected to be only a (linear) function of temperature, quite
independent of the nature of the particle.

Let us denote the ratio of the fermionic chemical potential to the
temperature by $u = \mu_F / T$. According to the above discussion
an ultra-relativistic gas of fermions will be fully described by
just two dimensionless parameters: the number of fermionic degrees
of freedom $f_F$ and the chemical potential per temperature $u$.

\subsection{The grand canonical formalism and some important abbreviations}

We now proceed to determine the thermodynamic properties of an
ultra-relativistic gas of fermions. The relevant quantity in the
grand-canonical formalism is the grand canonical potential $J(T,
\mu, V)$, which is defined as a function of temperature $T$,
chemical potential $\mu$ and volume $V$. For an ultra-relativistic
gas with $f_F$ fermionic degrees of freedom with energy-momentum
relation $\epsilon=p$ the grand canonical potential is given by:

\begin{equation} \label{eq:J:ln}
J(T, \mu, V) = - \frac{f_F}{(2 \pi \hbar)^3} T \, V \int \int \int
d^3p \, \ln{(1+e^{-\frac{p- \mu}{T}})}
\end{equation}

By introducing the dimensionless integration variable $z = p/T$ we
can cast $J$ into another form:

\begin{equation} \label{eq:J0}
J = - T^4 V \frac{f_F}{2 \pi^2 \hbar^3} \int_0^\infty{z^2
\ln{(1+e^{-z+u})}dz}
\end{equation}

where we have set the integration ranges to zero and infinity and
replaced the chemical potential per temperature with the
dimensionless parameter $u$:

\begin{equation} \label{eq:u}
u = \frac{\mu}{T}
\end{equation}

$u$ depends on $\mu$ and $T$, which are both {\em independent}
variables in the grand canonical formalism. Therefore, whenever we
calculate the thermodynamic quantities from the grand-canonical
potential via partial derivatives, we {\em must} treat $u$ as a
function of the {\em independent} variables $\mu$ and $T$.

The integral in equation (\ref{eq:J0}) can be transformed to the
following integral by a partial integration:

\begin{equation} \label{eq:J}
J = - T^4 V \frac{f_F}{2 \pi^2 \hbar^3} \frac{1}{3}
\int_0^\infty{z^3 n_F(z,u)dz}
\end{equation}

where $n_F$ is the mean occupancy number of the fermions:

\begin{equation} \label{eq:nF}
n_F(z, u) = \frac{1}{e^{z-u}+1} = \frac{1}{e^{\frac{p-\mu}{T}}+1}
\end{equation}

The thermodynamic equations for an ideal gas at ultra-relativistic
temperature can be expressed exclusively in terms of definite
integrals, whose integrand contains $n_F$ multiplied by an integer
power of $z$. I will denote these integrals by $Z_{F,n}$:

\begin{equation} \label{eq:ZF}
Z_{F,n}(u) = \int_0^\infty{z^n n_F(z,u)dz}
\end{equation}

Such integrals can be evaluated by the poly-logarithmic function
(see the Appendix for specific formula).

\subsection{Extrinsic quantities and densities}

According to the grand canonical formalism the entropy $S$ can be
calculated by a partial differentiation with respect to $J$:

\begin{equation} \label{eq:S:0}
S = -\frac{\partial J}{\partial T} = \frac{f_F}{2 \pi^2 \hbar^3} T^3
V \left(\frac{4}{3} Z_{F,3}(u) - u Z_{F,2}(u)\right)
\end{equation}

For the derivation of the
entropy the following identity has been used:

\begin{equation}
\frac{\partial {Z_{F,3}(u)}}{\partial x} = 3 Z_{F,2}(u)
\frac{\partial u}{\partial x}
\end{equation}

The pressure is given by:

\begin{equation} \label{eq:P}
P = -\frac{\partial J}{\partial V} = \frac{f_F}{2 \pi^2 \hbar^3}
\frac{Z_{F,3}(u)}{3} T^4 = \frac{F_E}{3} T^4
\end{equation}

where the following abbreviation was used:

\begin{equation} \label{eq:FE:0}
F_E = \frac{f_F Z_{F,3}(u)}{2 \pi^2 \hbar^3}
\end{equation}

With the above relation, the grand canonical potential $J$ can be
expressed in terms of $P$ and $V$:

\begin{equation} \label{eq:J:PV}
J = - \frac{F_E}{3} V T^4 = - P V
\end{equation}

The entropy $S$ can be expressed as:

\begin{equation} \label{eq:S}
S = F_S T^3 V
\end{equation}

with

\begin{equation} \label{eq:FS:0}
F_S =  \frac{f_F \left(\frac{4}{3} Z_{F,3}(u) - u
Z_{F,2}(u)\right)}{2 \pi^2 \hbar^3}
\end{equation}

The total energy is calculated from the grand canonical potential
via:

\begin{equation} \label{eq:EF}
E = J -\left(T \frac{\partial}{\partial T} + \mu
\frac{\partial}{\partial \mu}\right)J = \frac{f_F}{2 \pi^2
\hbar^3} T^4 V Z_{F,3}(u) = F_E T^4 V
\end{equation}

As expected, we find the equation of state for an ideal
ultra-relativistic gas:

$$e = \frac{E}{V} = 3 P$$

Throughout this paper the extrinsic quantities, such as total
energy $E$, total entropy $S$ etc. of a space-time region $V$ will
be denoted by capital letters, whereas the densities will be
denoted by lower case letters. Quantities referring to the properties
of the individual particles, such as the mean energy per particle
or the entropy per particle will be denoted by (lower case) greek
letters. With this notation the total energy is denoted by $E$,
the energy-density by $e$ and the energy per particle by
$\epsilon$.

The grand-canonical potential $J$ is related to the total energy
via the well-known relation:

$$J = -\frac{E}{3}$$

Keep in mind that $J$ is defined as a function of $T$, $V$ and
$\mu$, so that taking a partial derivative with respect to $E$ is
tricky.

The energy-density $e = E/V$ is proportional to the fourth
power of the temperature and proportional to the number of
ultra-relativistic degrees of freedom, via $f_F$:

$$e = \frac{E}{V} = F_E T^4$$

Combining equations (\ref{eq:S}, \ref{eq:EF}) we can derive a
relation between the entropy, the energy and the temperature:

\begin{equation} \label{eq:ST:kE}
S T = \frac{F_S}{F_E} E = \kappa E
\end{equation}

The ratio of $F_S/F_E$ will turn out important later, so we have
denoted it by $\kappa$:

\begin{equation} \label{eq:kappa}
\kappa = \frac{S T}{E} = \frac{F_S}{F_E}
\end{equation}

Note that according to equation (\ref{eq:kappa}) $\kappa$ is
defined exclusively in terms of the thermodynamic quantities $S$,
$T$ and $E$. This definition of $\kappa$ is completely general. It
appears, as if $\kappa$ can take on any (non-zero) value. However,
for most systems $\kappa$ is of order unity and nearly constant.
Take for example an ideal gas of {\em massive} particles at low
temperature. For $m \gg T$ the entropy per massive particle is
$\sigma \approx m/T$ and the total energy per particle is
$\epsilon = m + 3/2 \, T$, so that $\kappa \rightarrow 1$. For any
Kerr-Newman black hole with given angular momentum and charge
$\kappa$ is constant, and lies in the range $0 < \kappa \leq 1/2$
(see section \ref{sec:black:holes}).

For an ultra-relativistic gas $\kappa = F_S / F_E$, which is of
order unity and nearly constant whenever the number of particle
degrees of freedom does not change. For a gas consisting
exclusively out of fermions $\kappa$ only depends on $u$, but not
on the number of fermionic degrees of freedom $f_F$. If the gas
has a non-zero contribution of bosons, $\kappa$ also depends -
albeit very moderately - on the ratio of bosonic to fermionic
degrees of freedom.

The total number of particles is given by:

\begin{equation} \label{eq:NF}
N = -\frac{\partial J}{\partial \mu}=\frac{f_F}{2 \pi^2 \hbar^3}
Z_{F,2}(u) T^3 V = F_N T^3 V
\end{equation}

where we have defined the quantity

\begin{equation} \label{eq:FN}
F_N = \frac{f_F Z_{F,2}(u)}{2 \pi^2 \hbar^3}
\end{equation}

Like $F_E$ and $F_S$, the value of $F_N$ only depends on the
chemical potential per temperature $u$ and on the number of
ultra-relativistic degrees of freedom $f$. When these quantities
are fixed, then $F_E$, $F_S$ and $F_N$ are constants.

The number-density $n = N/V$ is proportional to the cube of the
temperature and proportional to the number of ultra-relativistic
degrees of freedom via $F_N$

\begin{equation} \label{eq:n}
n = \frac{N}{V} = F_N T^3
\end{equation}

We can combine equation (\ref{eq:n}) with equation (\ref{eq:P}) in
order to obtain a relation, which resembles the ideal gas law:

\begin{equation} \label{eq:gaslaw}
P V = \frac{F_E}{3 F_N} N T = R N T
\end{equation}

with

\begin{equation} \label{eq:R:gaslaw}
R = \frac{F_E}{3 F_N}
\end{equation}

In the non-relativistic case $R=1$ (in units $c=k=1$), as will be
shown in the Appendix. In the ultra-relativistic case $R$ depends
on the ratio of bosonic to fermionic degrees of freedom and on
$u$. Yet $R$ remains close to unity for reasonable assumptions
with respect to the values of $r = f_F/f_B$ and $u$. For $u=0$ we
get $R_B = \pi^4 / (90 \zeta(3)) \simeq 0.90039$ for bosons. For
fermions the value is higher by the factor $7/6$: $R_F = 7/6 \,
R_B = 7 \pi^4 / (540 \zeta(3)) \simeq 1.05046$.

\subsection{Thermodynamic parameters of individual particles}

In this section the thermodynamic parameters which refer to the
properties of the individual particles will be derived.

The energy per particle is proportional to $T$, as can be
seen by combining equations (\ref{eq:EF}, \ref{eq:NF}):

\begin{equation} \label{eq:EF_NF}
\epsilon = \frac{E}{N} = \frac{Z_{F,3}(u)}{Z_{F,2}(u)} \, \, T =
 \frac{F_E}{F_N} \, \, T
\end{equation}

The energy per particle depends linearly on the temperature, quite
as expected for an ultra-relativistic gas. The constant of
proportionality $F_E/F_N$ doesn't depend on the number of degrees
of freedom (for a gas consisting exclusively out of fermions). It
only depends on $u$, the ratio of the chemical potential to the
temperature. This ratio might depend indirectly on the
temperature. However, as has already been discussed in section
\ref{sec:properties:gas} and as we will see later, $u$ is
effectively constant when the number of ultra-relativistic degrees
of freedom of the different particle species making up the gas
does not change.

The entropy per particle $\sigma$ can be read off from equations
(\ref{eq:S}, \ref{eq:FS:0}, \ref{eq:NF}):

\begin{equation} \label{eq:SF_NF}
\sigma = \frac{S}{N} = \frac{4}{3} \frac{Z_{F,3}(u)}{Z_{F,2}(u)} - u = \frac{F_S}{F_N}
\end{equation}

Similar to the ratio $\epsilon / T = F_E/F_N$ the entropy per
particle $\sigma = F_S / F_N$ only depends on $u$.

$\sigma$ might be (slightly) temperature-dependent via $u$. Note,
however, that for a "normal" ultra-relativistic gas with non-zero
chemical potential we know, that the entropy per particle is
constant. For example, a photon gas has $\sigma_B \simeq 3.6$ and
a gas of massless fermions with zero chemical potential has
$\sigma_F \simeq 4.2$. Therefore it seems reasonable to assume,
that $\sigma$ is nearly constant even in the more general case,
where the chemical potential of the fermions is non-zero.

The nearly constant energy per particle per temperature $\epsilon
/ T$, and the entropy per particle $\sigma$ are related. We find:

\begin{equation}
\sigma = \frac{F_S}{F_E} \frac{\epsilon}{T} = \kappa \frac{\epsilon}{T}
\end{equation}

An interesting case is $\kappa = 1$. In such a case the (mean)
energy per particle per temperature $\epsilon / T$ is exactly
equal to the (mean) entropy per particle $\sigma$. For this particular
case the free energy $F$ is exactly zero. $F$ is defined as:

$$F = E - S T$$

If we divide this equation by the particle number $N$, we get the
free energy per particle $\phi = F/N$:

$$\phi = \epsilon - \sigma T = (1 -\kappa) \epsilon$$

The free energy-density $f = F/V$ follows from the above equation
by multiplication with $N/V$:

$$f = e - s T = (1 -\kappa) e$$

\subsection{\label{sec:extension:bosons}Extending the model for bosons}

The calculations have been carried through for fermions.
The equations for an ultra-relativistic boson gas are very similar
to the above equations. We have to replace:

\begin{equation} \label{eq:Replace:nB}
n_F(z,u) \rightarrow n_B(z,u) = \frac{1}{e^{z-u}-1}
\end{equation}

\begin{equation} \label{eq:Replace:zB}
Z_{F,n} \rightarrow Z_{B,n} = \int_0^\infty{z^n n_B(z,u) dz}
\end{equation}

For an ideal gas the individual contributions to the extrinsic
quantities can be summed up.

\subsubsection{\label{sec:three:parameters}The three fundamental parameters of an ultra-relativistic ideal gas: $f_F$, $f_B$ and $u$}

Before we put the bosonic and fermionic contributions together,
let us reflect on the fundamental characteristics that will
describe the most general case of an ultra-relativistic gas.

Although we won't be able to distinguish the ultra-relativistic
particles by their rest-mass or by the transverse components of
their spins, we still should be able to distinguish bosons from
fermions. We should also be able to count the different helicity
states of a particle (one for a neutrino, two for an electron
according to the Standard Model of particle physics). Furthermore,
it should be possible to discern the particles in an operational
sense, if they have different chemical potentials.

Therefore the only relevant thermodynamic characteristics of a gas
consisting of ultra-relativistic fermions and bosons appear to be
the respective degrees of freedom of fermions and bosons and their
respective chemical potentials. Let us denote the fermionic
degrees of freedom by $f_F$ and the bosonic degrees of freedom by
$f_B$.

Generally, i.e. at low energies, the different particle species
can have very differing values for the chemical potentials. It is
usually assumed that the chemical potential of a non-relativistic
particle is related to its rest-mass. There are some restraints.
Ultra-relativistic Bosons cannot have a positive chemical
potential\footnote{As can be seen in the Appendix, {\em
non-relativistic} bosons can have a positive chemical potential,
albeit not arbitrarily large. The maximum possible value for
$\mu_B$ is given by the particle's mass $\mu_B \leq m$. This shows
that in general any boson which is its own anti-particle, must
have a chemical potential of zero.}, as $Z_{B,n}(u)$ is a complex
number for positive $u$. Photons and gravitons, in fact all
massless gauge-bosons, have a chemical potential of zero, which
reflects the fact, that they can be created and destroyed without
being restrained by a particle-number conservation law.

Here we are considering a gas of ultra-relativistic particles,
where particle-antiparticle pair production will take place
abundantly. There will not only be particles around, but every
particle will come with its antiparticle. The chemical potentials
of a particle and its anti-particle add up to zero: $\mu +
\overline{\mu} = 0$. This restricts the chemical potential of the
bosons: As ultra-relativistic bosons cannot have a positive
chemical potential, the {\em chemical potential of any
ultra-relativistic bosonic species must be zero}, i.e. $\mu_B =
\overline{\mu_B} = 0$, whenever the energy is high enough to
create boson/anti-boson pairs. This restriction does not apply to
the fermions, which can have a non-zero chemical potential at
ultra-relativistic energies, as both signs of the chemical
potential are allowed. So for ultra-relativistic fermions we can
fulfill the relation $\mu_F + \overline{\mu_F} = 0$ with non-zero
$\mu_F$.

In the following discussion let us take the convention, that any
fermion with $\mu_F >0$ is classified as ordinary matter, so that
all of the anti-fermions have a negative chemical potential. As
has been discussed before, it is reasonable to assume that at
ultra-high temperatures all fermions will have the same universal
value for the chemical potential, which is expected to be
proportional to the temperature.\footnote{The constant of
proportionality could be zero, though.}

If the fermions have a non-zero chemical potential, we
can distinguish bosons from fermions in an operational
sense by their different chemical potentials. We don't have to
determine the particle's spin (integer for bosons, half-integer
for fermions).

According to the previous discussion an ideal (uncharged, locally
isotropic) gas at ultra-relativistic temperatures is characterized
by just three dimensionless numbers: The bosonic and fermionic
degrees of freedom $f_B$ and $f_F$ and the ratio of the chemical
potential of the fermions to the temperature, $u = \mu_F / T$.
Henceforth we will express the thermodynamic properties of the
system in terms of these three fundamental parameters.

For many of the following relations it is convenient to replace
$f_F$ and $f_B$ with the ratio of bosonic to fermionic degrees of
freedom:

\begin{equation} \label{eq:r}
r_f = \frac{f_B}{f_F}
\end{equation}

and the total number of degrees of freedom

\begin{equation} \label{eq:f:total}
f = 2 (f_F+ f_B)
\end{equation}

We take the convention here, that $f_F$ and $f_B$ denote the
degrees of freedom of one particle species, {\em including}
particle and antiparticle. With this convention a photon gas
($g=2$) is described by $f_B = 1$ (There are two photon degrees of
freedom: The photon is its own anti-particle; it comes in two
helicity states). All other particle characteristics, such as the
different chirality states for a Dirac-electron, are counted
extra. The total number of the degrees of freedom in the gas,
counting particles and anti-particles separately, will thus be
given by $f = 2(f_F + f_B)$.

\subsubsection{\label{sec:extension:general}Extending the thermodynamic relations to the general case}

We can use all the relations derived for a fermion gas simply by making
the following replacements:

\begin{equation} \label{eq:FE}
F_E \rightarrow F_E(u, f_F, f_B) = \frac{f_F \left( Z_{F,3}(u) +
Z_{F,3}(-u)\right) + 2 f_B Z_{B,3}(0)} {2 \pi^2 \hbar^3}
\end{equation}

\begin{equation} \label{eq:FN:full}
F_N \rightarrow F_N(u, f_F, f_B) = \frac{f_F \left( Z_{F,2}(u) +
Z_{F,2}(-u)\right) + 2 f_B Z_{B,3}(0)} {2 \pi^2 \hbar^3}
\end{equation}

\begin{equation} \label{eq:FS}
F_S \rightarrow F_S(u, f_F, f_B) = \frac{f_F\left(\frac{4}{3} \{Z_{F,3}(u)\} - u
[Z_{F,2}(u)]\right) + 2 f_B \frac{4}{3} Z_{B,3}(0)} {2 \pi^2 \hbar^3}
\end{equation}

For brevity commutator anti-commutator notation was (mis)used:

$$\{Z_{F,n}(u)\} = Z_{F,n}(u) + Z_{F,n}(-u)$$

and

$$[Z_{F,n}(u)] = Z_{F,n}(u) - Z_{F,n}(-u)$$

Once in a while we might want to determine the entropy-,
number- and energy-densities of the individual components of the
gas, i.e. for a single bosonic or fermionic degree of freedom.
In such a case we just have to set $f_B$ or $(f_F, u)$ to zero in
the in above defined quantities $F_E$, $F_N$ and $F_S$. For
example, in order to obtain the bosonic contributions we set $f_F
= u = 0$; in order to obtain the fermionic contribution (including
the fermionic antiparticles) we set $f_B = 0$. This procedure gives

\begin{equation}
N_B = F_N(0, 0, f_B) \, \, V T^3 = \frac{2 f_B Z_{B,2}(0)}{2 \pi^2
\hbar^3} \, \, V T^3 = f_B \frac{2 \zeta(3)}{\pi^2 \hbar^3} \, \,
V T^3
\end{equation}

for the number of bosons and

\begin{equation}
N_F+\overline{N_F} = F_N(u, f_F, 0) \, \, V T^3 = \{F_N\} \, \, V T^3
\end{equation}

for the number of fermions + anti-fermions, where we have defined
the quantity

\begin{equation} \label{eq:FN:ac}
 \{F_N\} = F_N(u, f_F, 0) = \frac{f_F \left( Z_{F,2}(u) +
Z_{F,2}(-u) \right)}{2 \pi^2 \hbar^3}
\end{equation}

in a somewhat abusive usage of anti-commutator notation.

Calculating the fermionic and anti-fermionic contributions
individually is a little bit more tricky. Here we must keep in
mind that the fermions are described by the terms with positive
$u$ and the anti-fermions with the corresponding negative value
$-u$. The number of anti-fermions is given by:

\begin{equation}
\overline{N_F} = \frac{f_F Z_{F,2}(-u)}{2 \pi^2 \hbar^3} \, \, V
T^3
\end{equation}

whereas the number of fermions is obtained by replacing $-u$ with
$u$ in the above formula.

The total entropy of the fermions is:

\begin{equation}
S_F = \frac{f_F \left(\frac{4}{3} Z_{F,3}(u) - u Z_{F,2}(u)
\right)} {2 \pi^2 \hbar^3} \, \, V T^3
\end{equation}

For the anti-fermions we have:

\begin{equation}
\overline{S_F} = \frac{f_F \left(\frac{4}{3} Z_{F,3}(-u) + u
Z_{F,2}(-u) \right)} {2 \pi^2 \hbar^3} \, \, V T^3
\end{equation}

For the calculation that follows we will not only need the sum of
the number of fermions and anti-fermions, but also their
difference:

\begin{equation} \label{eq:Delta:NF}
\Delta N_F = N_F - \overline{N_F} = [F_N] \, V T^3
\end{equation}

where again commutator-notation was (ab)used:

\begin{equation} \label{eq:FNN}
[F_N] =  \frac{f_F \left( Z_{F,2}(u) - Z_{F,2}(-u) \right) }{2
\pi^2 \hbar^3}
\end{equation}

The quantity just defined in equation (\ref{eq:FNN}) allows us to
express $F_S$ via $[F_N]$ and $F_E$:

\begin{equation} \label{eq:FS:FE:u:FN}
F_S = \frac{4}{3} F_E - u [F_N]
\end{equation}

Using the identities known for the poly-log function it is rather
easy to show that

\begin{equation} \label{eq:z2z2}
Z_{F,2}(u) - Z_{F,2}(-u) = \frac{u}{3} \left( \pi^2 + u^2 \right)
\end{equation}

so that

\begin{equation} \label{eq:FNN:u}
[F_N] =  \frac{f_F}{6 \hbar^3} \left(1 + \frac{u^2}{\pi^2} \right) u
\end{equation}

This allows us to express $u$ as an implicit function of $\Delta
n_F$, $T$ and $f_F$:

\begin{equation} \label{eq:u:implicit:TVN}
\frac{\Delta n_F}{T^3} =  \frac{f_F}{6 \hbar^3} \left(1 +
\frac{u^2}{\pi^2} \right) u
\end{equation}

Another important expression that can be simplified by the known
identities for the poly-log function is

\begin{equation} \label{eq:z3z3}
\frac{Z_{F,3}(u) + Z_{F,3}(-u)}{2 Z_{B,3}(0)} + 1 = \frac{15}{8}
\left( 1 + \frac{u^2}{\pi^2} \right)^2
\end{equation}

so that $F_E$ can be expressed in a closed form as a function of
$u$:

\begin{equation} \label{eq:FE:u}
F_E = \frac{f_F \pi^2}{15 \hbar^3} \left( \frac{15}{8} \left( 1 +
\frac{u^2}{\pi^2} \right)^2 + (r_f-1)\right)
\end{equation}

$F_S$ can be determined as an explicit function of $u$ via
equation (\ref{eq:FS:FE:u:FN}).

With relations (\ref{eq:FNN:u}, \ref{eq:FE:u}) one can determine
$u$ in a closed form. Rearranging equation (\ref{eq:FS:FE:u:FN})
we find

\begin{equation}
\frac{4}{3} - \kappa = \frac{u [F_N]}{F_E}
\end{equation}

Inserting the expressions for $[F_N]$ and $F_E$ into the above
formula we finally get

\begin{equation} \label{eq:u2/pi2}
1 - \frac{3}{4} \kappa =  \frac{\frac{u^2}{\pi^2} \left(1 +
\frac{u^2}{\pi^2} \right) }{\left(1 + \frac{u^2}{\pi^2} \right)^2
+ \frac{8}{15}(r_f-1)}
\end{equation}

Equation (\ref{eq:u2/pi2}) is a quadratic equation in the variable
$u^2 / \pi^2$ which is easy to solve. A closed formula will be
derived below. It is clear from the above relation that there is
only a (real) solution for $u$ if $\kappa \leq 4/3$, because the
right hand side of equation (\ref{eq:u2/pi2}) is always positive.
For $\kappa = 4/3$ the only possible value for $u$ is zero,
independent of the value of $r_f$. If $u$ is a solution, so its
negative value $-u$ is a solution as well. Therefore any non-zero
value of $u$ allows us to distinguish particles and anti-particles
by their respective positive / negative values of $u$.

\subsubsection{Fermionic weighting factors}

Non-zero $u$ is only possible for fermions. Ultra-relativistic
bosons always have $u=0$. Therefore the extrinsic thermodynamic
quantities of the bosons, such as number-, energy- and
entropy-densities can be calculated by multiplying the values
derived from the well known Planck-distribution with the number of
bosonic degrees of freedom $f_B$. For a gas of massless fermions
with zero chemical potential it is common practice to multiply the
fermionic degrees of freedom with the so called "fermionic
weighting factors", which relate the number-, energy-,
entropy-densities of a single fermionic degree of freedom to a
single bosonic degree of freedom. It is convenient to extend this
procedure for non-zero $u$. The weighting factor for the energy
density is given by

\begin{equation}
w_E(u) = \frac{Z_{F,3}(u)}{Z_{B,3}(0)}
\end{equation}

The weighting factor for the number-density is

\begin{equation}
w_N(u) = \frac{Z_{F,2}(u)}{Z_{B,2}(0)}
\end{equation}

The weighting factor for the entropy-density can be calculated
from the two other weighting factors

\begin{equation}
w_S(u) = \frac{\frac{4}{3} Z_{F,3} - u Z_{F,2}(u)}{\frac{4}{3}
Z_{B,3}(0)} = w_E(u) - u \, w_N(u) \frac{45 \zeta(3)}{2 \pi^4}
\end{equation}

To get the weighting factors for the anti-fermions, which will be
denoted by barred quantities, we just have to replace $u$ with
$-u$, i.e. $\overline{w_E} = w_E(-u)$ and $\overline{w_N} =
w_N(-u)$. For $u=0$ we get the well known factors $7/8$ and $3/4$
by which the energy-and number-densities of a gas of fermions
differ from those of a photon gas with the same number of degrees
of freedom.

\subsubsection{On the relation between the thermodynamic parameters $u$, $r_f$ and $\kappa$.}

From the relations given in the previous two sections it is clear,
that all thermodynamical quantities can be calculated in closed
form, whenever $u$ is known. $u$ depends implicitly on $r_f$ and
$\kappa$, as can be seen by inspection of equation
(\ref{eq:u2/pi2}).

\begin{figure}[ht]
\begin{center}
\includegraphics[width=12cm, bb = 32 431 561 809]{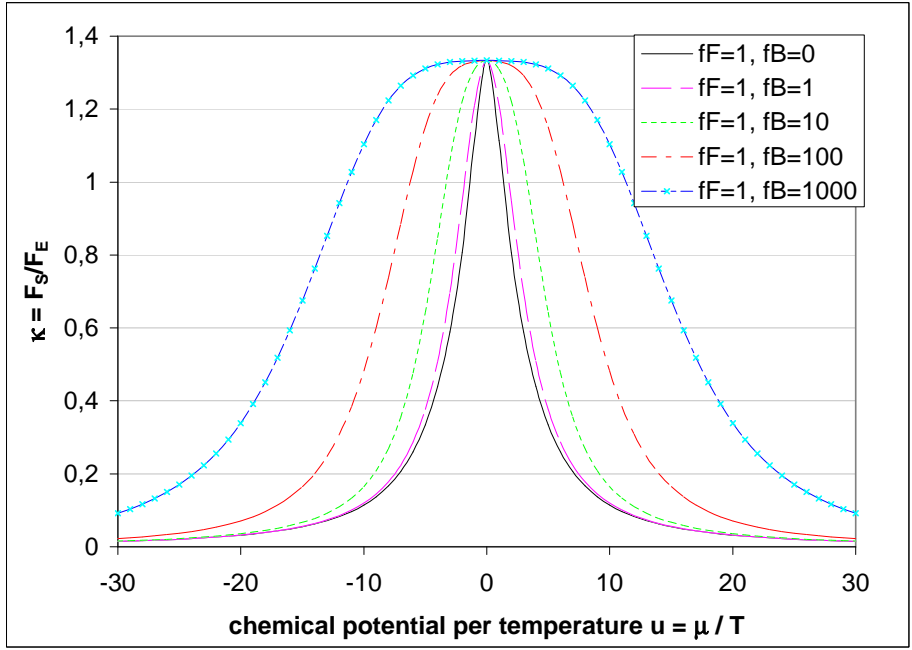}
\caption{\label{fig:kappa}$\kappa = F_S/F_E$ as a function of the
chemical potential per temperature of the fermions $u$, plotted
for various values of the ratio of bosonic to fermionic degrees of
freedom $r_f = f_B/f_F$}
\end{center}
\end{figure}

In order to get a better feeling of the functional relation
between $u$, $r_f$ and $\kappa$ it is instructive to plot
$\kappa(u, r_f)$ as a function of the (fermionic) chemical
potential per temperature $u$ for various - fixed - ratios $r_f =
f_B/f_F$. Figure \ref{fig:kappa} shows such a plot. $\kappa(u)$ is
a strictly positive, symmetric function of $u$ with a bell-shaped
form, similar to a Lorentz-profile. $\kappa(u)$ attains its
maximum value at $u=0$ and approaches zero quite rapidly for large
absolute values of $|u|$. The maximum value $\kappa(0) = 4/3$ is
independent of the parameter $r_f$, whereas the width of the
bell-shaped curve grows monotonically with $r_f$.

If $\kappa > 4/3$ there is no solution for $u$ in the equation
$\kappa(u, r_f) = const$, whatever the ratio of the number of
degrees of bosonic to fermionic degrees of freedom might be.
Negative values of $\kappa$ are not possible either. $\kappa = 0$
requires $u = \infty$, which doesn't seem to make sense from a
physical perspective. It is already clear from equation
(\ref{eq:kappa}) that $\kappa$ must be a positive quantity for any
reasonable closed thermodynamic system: $\kappa = S T / E$ can
only become negative if either the entropy, the temperature or the
total energy becomes negative. The microscopic statistical entropy
is always non-negative, negative temperatures only arise for
certain sub-systems and the total energy of a system is always
positive (at least in general relativity). $\kappa = 0$ requires
either zero temperature or entropy or infinite energy, which is
not physically sensible.

In the range $0 < \kappa < 4/3$ the chemical potential per
temperature is always non-zero, and depends on the ratio $r_f$.
For any given $r_f$ and $\kappa$ the corresponding value for $u$
can be read off from Figure \ref{fig:kappa} by determining the
intersection of the horizonal line $\kappa = const$ with the
bell-shaped curve parameterized by $r_f$. The innermost curve with
$r_f=0$ describes a gas consisting exclusively out of fermions.
Curves with larger $r_f$ lie above curves with lower $r_f$. All
curves have one point in common: The global maximum at $u=0$ with
$\kappa = 4/3$. From this construction it is clear that for any
given $0 < \kappa < 4/3$ the chemical potential per temperature
$u$ attains its minimum value for the innermost curve
parameterized by $r_f=0$ and that $u$ increases monotonically with
$r_f$ (for fixed $\kappa$).

\subsubsection{A closed formula for the chemical potential per
temperature of the fermions $u$}

Figure \ref{fig:kappa} allows a graphical determination of $u$. In
general one has to solve equation (\ref{eq:u2/pi2}). This relation
can be expressed in the following form, which allows an {\em
experimental} determination of $u$, whenever the ratios of
fermionic to bosonic number- and energy-densities is known.

\begin{equation} \label{eq:u:explicit}
u = (\frac{4}{3} - \kappa) \frac{\pi^4}{30 \zeta(3)} \frac{w_E +
\overline{w_E} + 2 r_f}{w_N - \overline{w_N}} = (\frac{4}{3} -
\kappa) \frac{\pi^4 }{30 \zeta(3) } \frac{\frac{e_F}{e_B}+
1}{\frac{\Delta n_F}{n_B}}
\end{equation}

where we have used

\begin{equation} \label{eq:wN:ratio}
\frac{w_N - \overline{w_N}}{2 r_f} = \frac{\Delta N_F}{N_B} =
\frac{\Delta n_F}{n_B}
\end{equation}

and

\begin{equation} \label{eq:wE:ratio}
\frac{w_E + \overline{w_E}}{2 r_f} = \frac{E_F}{E_B} =
\frac{e_F}{e_B}
\end{equation}

Note that $n_B$ is the (total) number-density of the bosons (all
bosons added up, with no distinction between bosons and
"anti-bosons"), whereas $\Delta n_F$ denotes the fermion number
density (number of fermions {\em minus} anti-fermions). $e_B$ and
$e_F$ denote the total energy-densities of bosons and fermions.

Using equation (\ref{eq:FNN:u}) one can express equation
(\ref{eq:u:explicit}) only in terms of the ratio $e_F / e_B$:

\begin{equation} \label{eq:u:explicit:2}
\frac{u^2}{\pi^2} (1 + \frac{u^2}{\pi^2}) = (\frac{4}{3} - \kappa)
\frac{2 r_f}{5} \left(\frac{e_F}{e_B}+ 1 \right)
\end{equation}

This is a quadratic equation in $u^2/\pi^2$, which is easy to
solve whenever $r_f$, $\kappa$ and the ratio of the
energy-densities of fermions to bosons is known.

Inserting equation (\ref{eq:z3z3}) into equation
(\ref{eq:u:explicit:2}) and using equation (\ref{eq:wE:ratio}) we
get a closed formula for $u$ in terms of the relevant parameters
$\kappa$ and $r_f$:

\begin{equation} \label{eq:u:closedform}
u = \pi \sqrt{\frac{2}{3 \kappa} \left(1 + \sqrt{1 + \frac{6}{5}
\kappa (\frac{4}{3}- \kappa)(r_f-1)}\right) - 1}
\end{equation}

For $\kappa = 4/3$ we find $u=0$, independent of $r_f$. Whenever
$u$ is small we can derive an approximation formula from equation
(\ref{eq:u:closedform}) or equation (\ref{eq:u:explicit:2}) for
small $u$:

\begin{equation} \label{eq:u:explicit:small}
u \approx \pi \sqrt{(\frac{4}{3}-\kappa)(r_f +
\frac{7}{8})\frac{2}{5}}
\end{equation}

\subsubsection{Supersymmetry}

In the important supersymmetric case $r_f=1$, i.e. equal number of
fermionic and bosonic degrees of freedom, equation
(\ref{eq:u:closedform}) is very much simplified:

\begin{equation} \label{eq:u:closedform:r=1}
u = \pi \sqrt{\frac{\frac{4}{3} - \kappa}{\kappa}}
\end{equation}

For $\kappa = 1$ we find

$$u = \frac{\pi}{\sqrt{3}}$$

so that

$$\frac{e_F}{e_B} = \frac{w_E + \overline{w_E}}{2} = \frac{7}{3} $$

The ratio of the entropy-densities of fermions to bosons turns out
to be simple, as well:

$$\frac{s_F}{s_B} = \frac{w_N \sigma_F + \overline{w_N \sigma_F} }{2 \sigma_B} = \frac{3}{2}$$

For $\kappa = 2/3$ we find ($r_f=1$)

$$u = \pi$$

so that

$$\frac{e_F}{e_B} = \frac{w_E + \overline{w_E}}{2} = \frac{13}{2} $$

and

$$\frac{s_F}{s_B} = \frac{w_N \sigma_F + \overline{w_N \sigma_F} }{2 \sigma_B} = \frac{11}{4}$$

and

$$\frac{\Delta n_F}{n_B} = \frac{w_N - \overline{w_N}}{2} = \frac{\pi^3}{6 \zeta(3)} \simeq 4.299$$

\subsubsection{Some more relations}

Generally it can be shown, that the ratio of the energy-,
entropy-and number-densities are related by

\begin{equation} \label{eq:sf/sb}
\frac{s_F}{s_B}  + 1= \frac{3 \kappa}{4} \left(\frac{e_F}{e_B} + 1
\right)
\end{equation}

and

\begin{equation}
\frac{\Delta n_F}{n_B} = (\frac{4}{3} - \kappa) \frac{ \pi^4 }{ 30
u \zeta(3) } \left( \frac{e_F}{e_B} + 1 \right)
\end{equation}

From equation (\ref{eq:sf/sb}) one can see, that for $\kappa =
4/3$ the ratios of the entropy-densities is equal to the ratio of
the energy-densities. This reflects the well known result for an
ultra-relativistic gas with zero chemical potential, where $s_F /
s_B = e_F / e_B = 7/8$.

An important quantity is the ratio of the energy-density of a
fermionic particle anti-particle pair to a bosonic particle pair:

\begin{equation}
\frac{w_E + \overline{w_E}}{2} = \frac{5 \omega^2}{6 \kappa^2} -1
\end{equation}

with

\begin{equation} \label{eq:omega}
\omega(\kappa, r_f) = 1 + \sqrt{1- \frac{6}{5} \kappa
(\frac{4}{3}- \kappa)(r_f-1) }
\end{equation}

For the supersymmetric case ($r_f = 1$) the above defined quantity
$\omega$  does not depend on $\kappa$. We have $\omega= 2$, so
that the ratio of the energy-densities reduces to

\begin{equation}
\frac{w_E + \overline{w_E}}{2} = \frac{10}{3 \kappa^2} -1
\end{equation}

The ratio of the number-density of a fermionic particle
anti-particle pair (number of fermions {\em minus} anti-fermions)
to a boson pair is given by:

\begin{equation}
\frac{w_N - \overline{w_N}}{2} = \frac{\pi^3}{18 \zeta(3)}
\frac{\omega}{\kappa} \sqrt{\frac{2 \omega}{3 \kappa}-1} = u
\frac{\pi^2}{18 \zeta(3)} \frac{\omega}{\kappa}
\end{equation}

where we have used:

\begin{equation}
u = \pi \sqrt{\frac{2 \omega}{3 \kappa} -1}
\end{equation}

In the supersymmetric case the ratio of the number densities reduces to

\begin{equation}
\frac{w_N - \overline{w_N}}{2} = \frac{\pi^3}{9 \zeta(3)}
\frac{1}{\kappa} \sqrt{\frac{4}{3 \kappa}-1} =  \frac{\pi^2}{9 \zeta(3)}
\frac{u}{\kappa}
\end{equation}

\subsection{\label{sec:free:energy}The free energy}

Knowing the grand canonical potential the free energy $F(T, V, N)$
can be calculated:

\begin{equation} \label{eq:F:def}
F = J + \sum{\mu N} = J + \mu_F N_F + \overline{\mu_F}
\overline{N_F} = -\frac{F_E}{3} V T^4 + u T \Delta N_F
\end{equation}

or somewhat differently, using equations (\ref{eq:Delta:NF}) and
(\ref{eq:FS:FE:u:FN}):

\begin{equation} \label{eq:F}
F = \left( u [F_N] -\frac{F_E}{3}\right) V T^4 = (F_E - F_S) V T^4
= E - S T
\end{equation}

The boson number $N_B$ does not show up in equations
(\ref{eq:F:def}) and (\ref{eq:F}) and the fermion number appears
as {\em difference} of the number of fermions minus anti-fermions.
This reflects the fact that the chemical potential of any
ultra-relativistic boson must be zero, whereas fermions are
expected to have a non-zero chemical potential.

In the {\em canonical} formalism the entropy is calculated from
the {\em free energy} by a partial differentiation with respect to
$T$. In order to do this, the free energy must be expressed as a
function of $T$, $V$ and the number of different particles $N_i$.
This is not the case with equation (\ref{eq:F}).

\subsubsection{\label{sec:fermion:conservation}A conservation law for the fermion number}

From equation (\ref{eq:F:def}) one can show that the fermion
number, i.e. the difference of fermion and anti-fermion numbers
$\Delta N_F = N_F - \overline{N_F}$, is a "conserved quantity"
whenever $u$ (or rather $\mu_F$) takes on a non-zero value. If we
take the partial derivative of $F$ with respect to $T$ we get

$$S = - \frac{\partial F}{\partial T} = \frac{4}{3} F_E T^3 V + \frac{1}{3} \frac{\partial F_E}{\partial T} T^4 V - \Delta N_F (u + T \frac{\partial u}{\partial T}) - u T \frac{\partial \Delta N_F}{\partial T}$$

$$= \frac{4}{3} F_E T^3 V - \Delta N_F u  + T\left(T^3 V [F_N] - \Delta N_F \right) \frac{\partial u}{\partial T}  - u T \frac{\partial \Delta N_F}{\partial T}$$

\begin{equation} \label{eq:S:from:F}
= \left( \frac{4}{3} F_E - u [F_N] \right) T^3 V - u T \frac{\partial \Delta N_F}{\partial T}
\end{equation}

where we have used

\begin{equation} \label{eq:partial:FE}
\frac{\partial F_E}{\partial T} = 3 [F_N] \frac{\partial u}{\partial T}
\end{equation}

and

\begin{equation}
\Delta N_F = [F_N] V T^3
\end{equation}

By comparing the result of equation (\ref{eq:S:from:F}) with
equation (\ref{eq:FS:FE:u:FN}) one can see, that the entropy
calculated from the free energy $F$ and the entropy derived from
$J$ ($S = F_S V T^3$) are only equal if

\begin{equation} \label{eq:delta:NF}
u T \frac{\partial \Delta N_F}{\partial T} = \mu \frac{\partial \Delta N_F}{\partial T} = 0
\end{equation}

This means that the total fermion number $\Delta N_F$ (=number of
fermions, counting fermions with a plus-sign and anti-fermions
with a minus-sign) must be independent of temperature, whenever
the chemical potential $\mu$ of the fermions is non-zero. It is
quite remarkable, that the assumption of non-zero chemical
potential gives us a {\em thermodynamic} "conservation-law" for
the total fermion number. At ultra-high temperatures there will be
several different fermionic species present, which will undergo
various particle-interchanging reactions. Yet the total fermion
number, i.e. the difference of fermions and anti-fermions is
conserved. Although the conservation of fermion number is built
into the Standard Model of particle physics (all of the
fundamental particles - excluding the gauge-bosons - are fermions
and can only be created in pairs), the Standard Model does not
really {\em explain} lepton or baryon number conservation. There
is no dynamical symmetry associated with the merely empirical
conservation of lepton ($L$) or baryon ($B$) number. Certain
supersymmetric theories only conserve $L-B$. In this respect it
comes somewhat of a surprise that there might be a thermodynamic
origin to the conservation of fermion number. This hints to some
deeper connection between thermodynamics, general relativity and
quantum field theory. Note also, that the thermodynamic equations
derived in the previous sections refer to the number-densities of
a boson particle anti-particle pair in a symmetrized version ($N_B
+ \overline{N_B}$ with $N_B = \overline{N_B}$), whereas the
fermion anti-fermion number-densities always appear as
anti-symmetrized quantities $\Delta N_F = N_F - \overline{N_F}$.

The fermion number "conservation law" can only be applied to
fermions, because - as discussed beforehand - only fermions {\em
can} have a non-zero chemical potential. For relativistic bosons
the chemical potential {\em must} be zero, therefore equation
(\ref{eq:delta:NF}) does not include the boson number
$N_B$.\footnote{For relativistic bosons it quite difficult, if not
impossible, to distinguish between "particles" and
"anti-particles", at least in a thermodynamic sense. The chemical
potential of the ultra-relativistic bosons is necessarily zero.
"Bosons" and "anti-bosons" are indistinguishable from the
viewpoint of thermodynamics. This reflects the current situation
in particle physics: It is well known, that all fundamental
massless bosons - such as the photon, the gluon and even the
hypothetical graviton - are their own anti-particles. (For
massless bosons one can distinguish the two helicity states, but
helicity is tied to $P$, not $C$ or $CP$.) Therefore it is
difficult to define a "boson number" in the same sense as a
fermion-, lepton- or baryon-number. $\Delta N_B = N_B -
\overline{N_B} = 0$. Therefore the only sensible way to calculate
the total boson number is given by $N_B + \overline{N_B}$ }. This
result is not quite unexpected: A conservation law for
boson-numbers would be in conflict with the fundamental physical
principle of gauge-invariance: All fundamental bosons are
gauge-bosons, i.e. they mediate the electro-magnetic, weak and
strong interactions between the (fermionic) particles of the
Standard Model. It is mandatory for a gauge boson that it can be
created (in a virtual process) without being restrained by a
particle-number conservation law.

\subsection{\label{sec:zero:chemical:potential}Zero chemical potential - a rather special case}

In this section I will discuss the rather special subcase of a
zero chemical potential of the fermions. We will see that a zero
chemical potential arises only under very special conditions.

The relevant thermodynamic quantities $F_S$ and $F_E$, which
appear in the relations for the energy- and entropy-densities of
an ultra-relativistic gas, only depend on $u$ and the number of
fermionic and bosonic degrees of freedom. At ultra-high
temperatures, above the unification scale, one expects that the
number of fundamental degrees of freedom does not change and that
the chemical potential per temperature remains constant. Therefore
it seems attractive to assume that $F_E$ and $F_S$ are independent
of temperature $T$. If this is actually the case, calculating the
entropy from the free energy via equation (\ref{eq:F}) is trivial.

In this rather special case one gets an independent expression for
the entropy, which - combined with equation (\ref{eq:S}) - allows
one to derive a relation between $F_E$ and $F_S$. This relation
fixes $\kappa$, so that we can determine $u$ whenever the number
of degrees of freedom $f_F$ and $f_B$ of the ultra-relativistic
gas is known.

There is no guarantee that $F_E$ and $F_S$, which depend on $f_F$,
$f_B$ and $u$ are independent of the thermodynamic variable $T$,
though. Even at ultra-high temperatures we cannot be sure that the
chemical potential per temperature $u$ of the fermions does not
depend on $T$. The number of ultra-relativistic fermionic and
bosonic degrees of freedom change whenever the temperature reaches
the mass-threshold of a particular species. This induces an
indirect dependence of $F_E$ and $F_S$ on temperature, which most
likely has an - indirect - effect on the value of $u$. Note also,
that the assumption of scale invariance at high temperatures not
necessarily requires $\mu \propto T$. A logarithmic dependence is
compatible with scale invariance as well. One therefore must check
very carefully if the results based on the assumption of
"constant" $F_E$ and $F_S$ are consistent. In fact, equation
(\ref{eq:partial:FE}) already tells us, that constant $F_E$ and
$F_S$ requires that $\partial u / \partial T = 0$ or $u=0$.

Assuming that $F_E$ and $F_S$ do not depend on $T$, one gets the
following alternative expression for the entropy:

\begin{equation} \label{eq:S:from:F:2}
-S = \frac{\partial F}{\partial T} = 4 (F_E - F_S) V T^3
\end{equation}

If we compare the above value for the entropy with the entropy
derived from the grand canonical potential, given by equation
(\ref{eq:S}), we can determine $\kappa$. We find $\kappa = 4/3$.

$\kappa$ has been expressed as a quadratic function of
$u^2$ and $r_f$ in equation (\ref{eq:u2/pi2}). A close look at
this equation shows (see also Figure \ref{fig:kappa}), that
whatever the value of $r_f$ might be, the only solution for
$\kappa = 4/3$ is $u=0$. Whenever an ultra-relativistic gas of
fermions and bosons can be described by $\kappa = 4/3$ the
chemical potential of the ultra-relativistic fermions must be
identical zero. This could already have been seen from equation
(\ref{eq:FS:FE:u:FN}).

The {\em assumption} that the chemical potential of the fermions
should be zero at high temperatures is not new. An argument for a
nearly zero chemical potential at high temperatures in a
cosmological context can be found in Weinberg's classical treatise
\cite[p. 531]{Weinberg/GR}. Weinberg's argument is based on the
assumption, that the number of photons in the universe at these
high temperatures is vastly larger than the number of baryons,
i.e. by a factor $10^9$.

The result $u=0$ is self-consistent. For $u=0$ the partial
derivative of $F_E$ and $F_S$ with respect to $T$ is zero,
trivially. Note also, that for $u=0$ the particle numbers $N_i$ of
the different species need not be considered in the expression for
the free energy (the chemical potentials of all particles are zero
and the $N_i$ are always multiplied by the chemical potentials
$\mu_i$ in the expression for the free energy). In the particular
case of an ultra-relativistic gas with zero chemical potential the
free energy $F$ is {\em equal} to the grand canonical potential
$J$, so that it is not surprising that it doesn't matter whether
we derive $S$ by a partial integration from $J = -(F_E/3) \, V
T^4$ or from $F = (F_E - F_S) \, V T^4$, assuming that $F_E$ and
$F_S$ are constant.

The combination $u=0, \kappa =4/3$ therefore is a perfectly
possible choice of parameters for an ultra-relativistic gas.
Furthermore, this choice minimizes the free energy to its {\em
least possible} value: $F = -E/3$. Why then consider the more
general case $\kappa < 4/3$ and $u \neq 0$, where the free energy
is higher?

It is not possible to give a short answer just right now. Exterior
constraints might impose a different value of $\kappa$ on the
whole system. We will see in the next section that self
gravitating systems are characterized by values of $\kappa$ which
lie in the range $0 < \kappa \leq 1$. Although gravity is a weak
force, it has the virtue of being positive all the time. For
sufficiently large systems it will be quite difficult to escape
the exterior constraints imposed on a thermodynamic system by the
general theory of relativity.

\section{\label{sec:self:gravitating:systems}On the relation between entropy, energy and free energy for self gravitating systems}

The purpose of this section is to shed some light on the physical
interpretation of the thermodynamic parameter $\kappa$, which
turned out to be of some relevance in the previous section. We
will see that whenever the entropy of a closed system can be
expressed as a function of the total energy alone, i.e. $S = S(E)$
(and when $\kappa \approx const$) the entropy $S$ and the total
energy $E$ are related by a power-law. $\kappa$ then is nothing
else than the exponent in the relation between $S$ and $E$. Some
examples for physical systems which can be described by such a
power relation are given at the end of this section.

\subsection{\label{sec:E:propto:F}A linear relation between the total and the free energy}

Equation (\ref{eq:ST:kE}) implies the following relation between
the free energy $F$ and the total energy $E$.

\begin{equation} \label{eq:F:kappa}
F = E - S T = (1-\kappa) E
\end{equation}

We find, that $F$ and $E$ are proportional to each other, with the
factor of proportionality given by $1-\kappa$. Note that the above
equation doesn't necessarily imply a strictly linear relation. Equation
(\ref{eq:ST:kE}) is completely general, as long as no assumptions
with respect to $\kappa$ are imposed. One can regard the
above equation as a definition for $\kappa$.

At ultra-relativistic temperatures we have seen that the allowed
range for $\kappa$ is quite restricted: $0 < \kappa \leq 4/3$. It
is clear that $\kappa$ can only vary very moderately within a
particular thermodynamic model, at least at high temperatures.
Here we are not interested in the most general thermodynamic model
possible, but rather in the cases which are relevant from a
physical perspective. As has been already discussed in the
previous section, at ultra-high temperatures one expects $u$ to be
nearly constant: Due to the lack of a specific mass- or energy
scale at ultra-high temperatures the chemical potential of an
ultra-relativistic fermion will depend nearly linearly on the
temperature\footnote{scale invariance also allows a logarithmic
dependence}. As $\kappa$ is a function of $u$ and $r_f$ only, it
must be nearly constant whenever the ratio of bosonic to fermionic
degrees of freedom $r_f$ does not change and the chemical
potential of the ultra-relativistic fermions has the expected
nearly linear temperature-dependence.

From equation (\ref{eq:F:kappa}) one sees, that any thermodynamic
system with constant $\kappa$ is characterized by the property,
that it's free energy $F$ is exactly proportional to the total
energy $E$. The value $\kappa = 1$ is special. It renders the free
energy exactly zero, regardless of the total energy of the system.

Another interesting value for $\kappa$ was pointed out in the
previous section: For $\kappa = 4/3$ the chemical potential of all
particles zero, regardless of the number of degrees of freedom of
fermions and bosons. In this case we attain the well-known result,
that the free energy $F$ of an ultra-relativistic gas (with zero
chemical potential) is equal to the grand-canonical potential $J$.
However, the free energy is negative with $F = J = -E/3$.

Are there other values of $\kappa$ which might be of relevance and
are there actually systems - in theory and in praxis - that can be
described by a specific value of $\kappa \neq 4/3$?

\subsection{\label{sec:S=S(E)}Thermodynamic systems with S = S(E)}

Before I attempt to answer this question, it is instructive to
find out what more can be inferred about the thermodynamic
properties of a closed system with nearly constant $\kappa$. In
order to make specific predictions some additional assumptions are
required. We are interested in the thermodynamic behavior of a gas
at ultra-high temperatures, i.e. temperatures well above the
electron mass threshold. Such extreme temperatures are only
conceivable for self-gravitating systems, such as in the early
phase of the universe or during relativistic collapse of a massive
star. For sufficiently compact gravitationally bound objects, such
as neutron stars (and even more so for black holes) there is a
more or less rigid relationship between size (= boundary area) and
the total energy (=asymptotic gravitating mass) of the system. It
appears as if this is a universal property of bounded self
gravitating systems, whose internal energy sources - such as
thermo-nuclear reactions - have ceased so that their energy output
is determined by gravitational phenomena alone.

Whenever there is a definite relationship between the system's
spatial extension and its total energy, i.e. $V = V(E)$ - or
alternatively $A = A(E)$, where $A = \partial V$ is the system's
boundary area - the temperature can only depend on $E$. This
follows from the ultra-relativistic gas law $E \propto V(E) T^4$
and the assumption of scale-invariance at ultra-high
temperatures\footnote{Scale-invariance is required in order to
ensure, that the constant of proportionality in the
ultra-relativistic gas law only depends - moderately! - on
temperature.}.  In this case the entropy will also be an exclusive
function of $E$, as $S \propto V T^3$ at ultra-relativistic
temperatures. The result is, that whenever the temperature of a
system is high enough and it's spatial extension depends only on
its total energy, the system's entropy can be expressed as a
function of $E$ alone:

$$S = S(E)$$

This is even true for the most extreme version of a
self-gravitating system, a black hole: Although the "volume" of a
black hole cannot be properly defined and it's interior matter
state clearly is not that of an ultra-relativistic gas, it's
boundary area $A$ remains a meaningful concept. For a spherically
symmetric black hole we find $S \propto A \propto E^2$. In the
more general case of a Kerr-Newman black hole the entropy also
depends on the exterior conserved quantities, which are the black
hole's angular momentum $J$ and charge $Q$. However, this
additional complication does not change the results that will be
obtained in this and the following sections. The essential
requirement is that the total differential of the entropy only
depends on {\em one of the three} variables $E, V, N$, i.e. $dS =
(\partial S / \partial E) dE$ + terms independent of $V$ and $N$.
This requirement is fulfilled by all classical black hole
solutions.\footnote{For all black hole solutions the number of
particles $N$ and the volume $V$ is undefined. Yet $A = \partial
V$ is still defined and there exists a one-to-one relationship $E
\leftrightarrow A$. Therefore the total energy $E$ and the area of
the event horizon $A$ are interchangeable. However, $N$ cannot be
expressed in terms of $E$ or $A$ for a black hole. In the
holographic solution all three quantities $E$, $V$ and $N$ are
interchangeable. Furthermore the total volume $V$ of a holostar is
well defined and is related one-to-one to its boundary area $A$.}

Whenever the entropy $S$ happens to be a function of the energy
$E$ alone, one can use the well known thermodynamic identity:

\begin{equation} \label{eq:SET}
\frac{\partial S}{\partial E} T = 1
\end{equation}

and replace the partial derivative with a total derivative. This
allows one to express the temperature as the total derivative of
$E$ with respect to $S$:

$$T = \frac{dE}{dS}$$

Inserting this into equation (\ref{eq:ST:kE}) we get:

\begin{equation} \label{eq:dSdT}
\frac{dE}{E} = \kappa \frac{dS}{S}
\end{equation}

This equation can be integrated, whenever $\kappa$ is a function
of energy (or entropy) alone. If $\kappa$ is constant, the
integral is trivial:

\begin{equation} \label{eq:S:E:kappa}
S = \alpha \cdot E^{\frac{1}{\kappa}}
\end{equation}

$\alpha$ is an integration constant. We find, that whenever the
entropy of a closed system is related to its total energy by a
power-law, $\kappa$ is nothing else than the exponent describing
this relation, i.e. $E \propto S^{\kappa}$. The temperature
follows from equation (\ref{eq:SET}):

\begin{equation} \label{eq:T:kappa}
T = \frac{\kappa}{\alpha} E^{1-\frac{1}{\kappa}}
\end{equation}

It is important to keep in mind, that the relation between total
energy and total entropy is a {\em global} one, which is only
valid for the system as a whole. The temperature derived in
equation (\ref{eq:T:kappa}) thus must be interpreted as the
"global" temperature of the system in question. A global relation
between entropy and energy does not necessarily mean, that one can
define a local temperature at every space-time point.\footnote{For
some systems this seems to work well, though. In the holostar
space-time the local temperature at any space-time point can be
derived consistently from the global temperature at infinity.
However, the specific value of $\kappa$ seems to depend on the
reference point from which $E$ is determined. This is not totally
unexpected: $\kappa$ is determined from $S$, $T$ and $E$. Whereas
the entropy $S$ is a pure number, independent of the local
reference point, the total energy of a gravitationally bound
system clearly depends on the reference point from which it is
calculated (for example: the energy of a particle in Schwarzschild
space-time starting out from infinity is always zero at the black
hole's horizon). In the holostar space-time one can calculate $E$
at spatial infinity ($\kappa = 1/2$), at the boundary ($\kappa =
3/4$), for a co-moving observer in the interior ($\kappa = 2/3$)
and for a stationary observer in the interior ($\kappa=1$). See
section \ref{sec:holo} for a more detailed discussion.}

All this has been quite abstract. The question that naturally
arises at this point is, whether we actually can find (closed)
systems in nature that are characterized by a power-law dependence
between $E$ and $S$, or more generally by a relation $S = S(E)$.
It appears, that $S = S(E)$ is a common occurrence among
self-gravitating systems.

\subsection{\label{sec:black:holes}Black hole solutions}

Let us first apply equations (\ref{eq:S:E:kappa},
\ref{eq:T:kappa}) to the simplest theoretical self gravitating
system known, a spherically symmetric black hole. For the
Schwarzschild black hole we have the (global) relation $S = 4 \pi
E^2 / \hbar$, where $E$ is taken to be the gravitational mass of
the black hole measured at spatial infinity.\footnote{We will work
in units $c=G=1$ throughout this and the following sections.} We
find $\kappa = 1/2$ and $\alpha = 4 \pi / \hbar$. Inserting these
values into equation (\ref{eq:T:kappa}) then gives us - not
unexpectedly - the right relation for the Hawking temperature of a
black hole, measured at spatial infinity, i.e. $T = \hbar / (8 \pi
E)$.

According to the relation $F = (1-\kappa) E$, the free energy for
a spherically symmetric black hole amounts to {\em half} it's total
energy. We will see later, that there are exact solutions to the
classical field equations, which minimize $F$ to zero in the
interior, matter-filled region. In a certain sense one can say,
that a classical black hole does a good job in minimizing the free
energy {\em globally} ($F = E/2$), but fails to minimize the free
energy {\em locally} to the least possible value conceivable for a
self-gravitating system ($F=0$).

The relation $S = S(E)$ can be extended to charged and rotating
black holes. Although the relation between the total energy and
entropy of the more general black hole solutions also depends on
the total charge(s) $Q_i$ and the total angular momentum $J$,
these quantities are globally {\em conserved}. The total
differential of the entropy $dS$ only depends on $dE$, $dQ_i$ and
$dJ$, but not on $dV$ or $dN$. For an isolated black hole in an
exterior vacuum space-time the conserved quantities cannot change,
so that $dJ = dQ_i = 0$. Whenever we can consider the black hole
and its surroundings as a closed system, we should not treat the
exterior conserved quantities as independent thermodynamic
variables, so that the temperature can be derived from a {\em
total} derivative $T = \partial E / \partial S = dE/dS$ whenever
$dJ = dQ_i = 0$.

This can be seen as follows: For a Kerr-Newmann black hole
$\kappa$ can be calculated from its definition in equation
(\ref{eq:kappa}). Using the well known relations for the mass
$M=E$, entropy $S$ and temperature $T$ of a Kerr-Newman black hole
we find:

\begin{equation} \label{eq:kappa:bh:M}
\kappa = \frac{S T}{E} = \frac{1}{2} \sqrt{1-\left(
\frac{Q}{M}\right)^2 - \left( \frac{J}{M^2}\right)^2 }
\end{equation}

or alternatively, as a function of $J, Q$ and area $A$:

\begin{equation} \label{eq:kappa:bh}
\kappa = \frac{1}{2} \frac{1 - \left(\frac{4 \pi Q^2}{A}\right)^2
- \left(\frac{8 \pi J}{A}\right)^2 }{ \left(1 + \frac{4 \pi
Q^2}{A}\right)^2  + \left(\frac{8 \pi J}{A}\right)^2 }
\end{equation}

$\kappa$ lies in the range $0 \leq \kappa \leq 1/2$ and only
depends on the dimensionless ratios $Q/M$ and $J/M^2$
(alternatively on $Q^2/A$ and $J/A$). For an extreme Kerr-Newman
black hole $\kappa = 0$. In this case the black hole's free energy
is equal to its total energy. This is the maximum possible value
for the free energy. For a general Kerr-Newman black hole of given
mass $M$ or area $A$ the free energy $F = (1-\kappa) E$ is
minimized, when $\kappa$ takes on its maximum value. This is the
case for $\kappa=1/2$, corresponding to $Q=J=0$, i.e. a
spherically symmetric Schwarzschild black hole. The expectation,
that the Schwarzschild black hole should be the thermodynamically
preferred state, which minimizes the free energy, is quite in
agreement with the general expectation, that any charged or
rotating black hole will shed angular momentum and charge, until
it becomes a spherically symmetric Schwarzschild black hole. A
very important parameter for the more general black hole solutions
is the irreducible mass, which is proportional to the black hole's
horizon area / entropy. Whereas the mass $M$ of a black hole can
change in the various allowed processes which extract energy (but
not entropy) from a black hole, a black hole's mass can never be
lowered below its irreducible mass by any classical process. For a
Schwarzschild black hole its irreducible mass is identical to its
gravitating mass, so - from the classical viewpoint - a further
reduction of the total energy is not possible.

It is possible to determine the correct relation between entropy
and total energy of a Kerr-Newman black hole through $\kappa$,
assuming that the partial derivative $1/T = \partial S / \partial
E$ can be replaced by a total derivative:

$$\kappa = \frac{S T}{E} = \frac{S}{E} \frac{\partial E}{\partial S} = \frac{S}{E} \frac{dE}{dS} $$

This is a differential relation between $S$ and $E$, which can be
integrated:

$$\frac{dS}{S} = \frac{dE}{\kappa E} = \frac{2 dE}{E \sqrt{1-\left( \frac{Q}{E}\right)^2 - \left( \frac{J}{E^2}\right)^2 }}$$

This integration is easy to perform by a change of variables $x =
E^2$:

\begin{equation}
\ln{S} = \int{\frac{dx}{\sqrt{x^2 - Q^2 x - J^2}}} = \ln{(2
\sqrt{x^2 - Q^2 x - J^2} + 2 x - Q^2)} + C
\end{equation}

Exponentiating the above result and replacing $x$ with $E^2 = M^2$
gives us the following relation:

\begin{equation}
S \propto 2 M^2 + 2 \sqrt{M^2 - Q^2 - \frac{J^2}{M^2}} - Q^2
\end{equation}

This is the correct formula for the entropy of a Kerr-Newman black
hole, if we multiply the right hand side by $\pi / \hbar$.

Note that the quadratic correspondence between energy and entropy
$S \propto E^2$ holds approximately for charged and rotating black
holes. For instance, an extreme Kerr black hole with $M^2 = J$
and $Q=0$ has $S = 2 \pi J / \hbar = 2 \pi M^2 / \hbar$. For
an extremely charged Reissner Nordstr\"{o}m black hole with $J=0$
and $Q=M$ we find $S = \pi Q^2 / \hbar = \pi M^2 / \hbar$.

\subsection{\label{sec:holo}The holographic solution}

Classical black holes are {\em vacuum} solutions of the field
equations and therefore not very well suited for a thermodynamic
analysis based on microscopic statistical thermodynamics, which
requires matter. Are there other systems to which equations
(\ref{eq:S:E:kappa}, \ref{eq:T:kappa}) apply?

Recently the so called holographic solution was discovered in
\cite{petri/bh}. The holographic solution is an exact solution of
the original Einstein field equations. It describes a
self-gravitating system of arbitrary size with an interior string
equation of state. The holostar has properties very similar to a
black hole. Most notably its entropy and temperature at infinity
are proportional to the Hawking result, as can be shown by simple
microscopic statistical thermodynamic analysis of the interior
matter-filled space-time. The holographic solution and its
properties are discussed extensively in \cite{petri/hol,
petri/thermo, petri/charge, petri/string}. Here I will just point
out some of the results which are relevant in the context of this
paper.

For any thermodynamic system the total energy $E$ is one of its
key variables. In contrast to the entropy, which is a pure number,
and the temperature, which is a {\em local} parameter, whose
normalization is tied to the total energy, the concept of total
energy is somewhat ambiguous in a curved space-time. In general
relativity it is of paramount importance to define the point of
reference, from which the total energy of the space-time is
calculated. In the black hole case the only sensible reference
"point" is the position of an asymptotic observer at spatial
infinity. The temperature and energy attributed to a black hole
refer to this point of reference. In contrast to a black hole the
holostar solution is regular throughout the whole space-time.
There is no event-horizon which separates two causally
disconnected regions. Therefore other choices of reference-point
are possible.

If we take spatial infinity as reference point, we get
$\kappa=1/2$, equal to the black hole case. This is not
unexpected, because a holostar viewed from the exterior space-time
is practically indistinguishable from a black hole.

However, another natural reference point is the position of the
holostar's spherical boundary membrane. The effective potential
for the geodesic motion of particles has a global minimum at this
position, so that the membrane might be viewed as a "better"
reference position for a stationary observer. If one calculates
the total energy with respect to this reference point (by a {\em
proper} integral over the interior energy-density) one gets $E
\propto r_h^{3/2}$, where $r_h$ is the radial coordinate position
of the holostar's boundary. For large holostars $r_h$ is nearly
identical to the holostar's gravitational radius, so that $r_h
\simeq 2 M$. The entropy of the holostar is proportional to its
boundary area, i.e. $S \propto r_h^2$. Therefore the holostar is
characterized by $S \propto E^{4/3}$, if the total energy is
evaluated at the boundary membrane. This dependence translates to
$\kappa = 3/4$, which predicts $T \propto E^{-1/3} \propto
r_h^{-1/2}$, which is the correct temperature dependence for the
holostar's surface temperature at the boundary membrane. Using the
exact numerical figures from \cite{petri/hol} (and setting $\hbar
=1$) one finds $\alpha = \pi \, 3^{4/3} r_0^{2/3}$, from which $T
= 1 / (4 \pi \sqrt{r_0 r})$ follows, exactly the result derived in
\cite{petri/hol} and independently in \cite{petri/thermo}. $r_0$
is a (constant) scale factor with dimensions of length, roughly
equal to twice the Planck length.\footnote{In \cite{petri/thermo}
the thermodynamic properties of the holostar solution were derived
by microscopic statistical thermodynamics, assuming that the
interior matter state can be described by a gas consisting of
ultra-relativistic particles. The entropy of the holostar scales
as $r^2$, it's internal temperature with $1/\sqrt{r}$ and the
temperature measured at infinity with $1/r$. Therefore the
holographic solution reproduces the Hawking result up to a
constant factor. However, the constant factor could not be
determined in \cite{petri/thermo}, because it depends on the total
number of particle degrees of freedom at high energies, which is
not known. Yet it is possible to equate the holostar's temperature
at infinity to the Hawking temperature, which then allows an
unambiguous determination of the interior temperature and the
entropy, and as a by-product an estimate for the total number of
fundamental particle degrees of freedom at ultra-high
temperatures.}

A third natural reference point is that of a {\em geodesically
moving} observer in the {\em interior} space-time. Note, that this
is a {\em local} point of reference, as the observer moves - in
highly relativistic motion - through the interior space-time. But
the local observer sees only a small fraction of the whole
space-time: He can never look beyond his current
Hubble-radius.\footnote{In the holostar space-time the local
Hubble-radius grows with time $r_H \propto 1/H = t$, similar to
the behavior of the Hubble-radius in an isotropically expanding
FRW-type universe. There is some - albeit quite tentative -
evidence, that Hubble-radius of the geodesically moving observer
in the interior holostar space-time might be identified with a
local acceleration horizon. A geodesically moving observer in the
interior holostar space-time is accelerated (as viewed from the
static coordinate frame). The proper acceleration falls off over
time. Furthermore, due to the negative radial pressure even a
(nearly) geodesically moving observer will feel a slight
deceleration in his frame, which falls off over time. The distance
to the acceleration horizon is inverse proportional to the
acceleration. This means, that the acceleration horizon $r_a$
grows with time. There is some evidence, that $r_a \propto t$.}
One cannot expect that the global relations are applicable to this
local observer. If we restrict the calculation of "total" entropy
and energy to the local Hubble-volume of the co-moving observer,
we get $E \propto r$ and $S \propto r^{3/2}$, where $r$ is the
current value of the scale factor ($H \propto 1/r$). This
translates to $\kappa = 2/3$, which predicts a temperature
dependence $T \propto 1/\sqrt{E} \propto 1 / \sqrt{r}$, again
exactly the temperature dependence that the co-moving observer
experiences in his local Minkowski frame.

Quite interestingly, the {\em local} thermodynamic properties of
the {\em interior} holostar space-time can also be characterized
by a different constant value of $\kappa$, which is tied to the
viewpoint of a - most likely hypothetic - {\em stationary}
observer at constant radial coordinate position $r$. In
\cite{petri/thermo} it has been shown, that the holostar's total
entropy and its temperature at infinity are exactly proportional
to the Hawking result if one calculates the local entropy-density
and temperature in the ($t, r, \theta, \varphi$)
coordinate-system, where the holostar space-time appears
static.\footnote{The total entropy is determined by proper
integration of over the local entropy density. The temperature at
infinity is the red-shifted surface-temperature.} However, unless
$\kappa = 1$ the holostar's temperature at infinity $T_\infty$
and its total entropy $S$ are not related with the right
factor.\footnote{The entropy and the temperature of a system are
conjugate variables. If one variable is normalized to a specific
value, the other variable follows unambiguously from the
thermodynamic identities. In the case of entropy and temperature
both variables are related via $\partial S / \partial E = 1/T$.
For the holostar solution it is fairly easy to show that $S
\propto E^2$ and $T_\infty \propto 1/E$, where $E =M$ is the
gravitational mass of the holostar measured by an observer at
spatial infinity. These dependencies are identical to the black
hole case. It was not possible to establish whether the holostar's
temperature and entropy are {\em identical} to the Hawking result
for a spherically symmetric black hole, because this would have
required knowledge of the total number of ultra-relativistic
degrees of freedom at the unification energy. This number is
unknown. It depends on the GUT-model. Furthermore, as the
GUT-scale is close to the string scale one expects a significant
string-contribution to the total matter state. What one can show
however is, that the thermodynamic identity $\partial S / \partial
E = 1/T$, relating entropy and temperature in the {\em exterior}
space-time, is only fulfilled if $\kappa = 1$ in the {\em
interior} space-time (or equivalently, if the free energy density
in the interior space-time is zero).}  If the holostar-solution is
to reproduce the Hawking-result with the correct relation between
$T$ and $S$, $\kappa$ must be unity in the {\em interior}
space-time. This result is quite remarkable, as $\kappa = 1$
implies that the free energy $F$ be identical zero. Therefore the
holostar minimizes it's {\em interior} free energy to zero, at
least from the viewpoint of a stationary interior
observer.\footnote{$\kappa = 1$ implies $T = const \approx T_{Pl}$
via equation (\ref{eq:T:kappa}), which is in conflict with $T
\propto 1 / \sqrt{r}$ in the holostar's interior space-time. One
possible solution is, that the relevant temperature for a
stationary observer is not the radiation temperature (seen by a
geodesically moving observer who passes the stationary observer),
but rather the temperature attributed to the local geodesic
acceleration that is required to keep the observer stationary,
i.e. his Unruh-temperature $T_U = a \hbar / (2 \pi)$. The geodesic
acceleration in the stationary frame is proportional to $a \propto
1/r^{3/2}$, which quite clearly is not constant. However, if one
transforms the proper geodesic acceleration to the frame of a
geodesically moving observer (with the implicit assumption, that
this is the preferred local Minkowski frame, to which all
measurements should be referenced), one finds that $\overline{a} =
const$, because the geodesically moving observer has a high
$\gamma$-factor $\gamma \propto \sqrt{r}$ and the proper
acceleration transforms with $\gamma^3$.}

\subsection{Friedmann-Robertson-Walker type solutions}

Does the relation $S \propto E^{1/\kappa}$ hold in other contexts?
There is another self-gravitating system, which we are quite
familiar with. It is called "the universe". Can the universe be
described by a relation between $S$ and $E$ with $\kappa = const$?

In order to answer this question one has to analyze the evolution
of energy- and entropy-densities in the universe's different
evolutional stages (matter-dominated, radiation-dominated, etc.).
Quite clearly this analysis is model-dependent. In this section
the analysis will be done in the context of the standard Friedmann
Robertson-Walker (FRW) model for the universe, which is based on
the so called cosmological principle. In the next section I will
discuss a somewhat more exotic model of a flat, holographic
universe.

In an homogeneously expanding space-time the expansion must be
nearly adiabatic, as there can be no net heat flow into or out of
a particular co-moving volume element in a homogeneous universe
(unless one assumes extra dimensions). With this assumption
entropy is approximately conserved in any co-moving volume. This
means, that the entropy-density $s$ will always be proportional to
the inverse volume. With the known relation $R \propto 1/T$
between the scale factor $R$ and the temperature $T$ we find $s
\propto T^3$.

In the matter-dominated era the energy-density $e$ has the same
behavior: $e \propto 1/R^3 \propto T^3$ (unless we assume that
particles are created or destroyed, violating local
energy-conservation in a co-moving volume\footnote{Particle/energy
creation is assumed in space-times governed by the perfect
cosmological principle, or in space-times with vacuum-energy.}).
Therefore $s \propto e$, which suggests $\kappa = 1$. However,
there is a slight difficulty in applying this finding to the
framework developed so far. We have assumed a power-law between
the {\em total} entropy and energy, which not necessarily
translates to the same power-law for the densities. Determining
the "total entropy" or the "total energy" of the universe is
tricky, if not impossible. We would have to multiply the
respective densities with the "total volume of the universe" at a
given time. This might not be well defined, particularly for an
open universe, which cannot be ruled out by today's measurements.

Yet there is an interesting observation: The product of the
entropy-density (which scales with $T^3$) and the "volume of the
universe" (which scales with $R^3 \propto 1/T^3$) is constant. The
same applies to the product of the energy-density with the volume
for the matter-dominated era. So we might still be able to relate
the total (constant) entropy to the total (constant) energy of the
universe, by calculating the (finite) ratio of the entropy to the
energy in any finite co-moving region. This ratio is time- and
position-independent for a matter-dominated FRW-universe. As the
ratio is the same for any arbitrary finite region, it is
reasonable to assume that $S \propto E$ for the whole universe,
even if its total volume, energy and entropy may be infinite. Note
also, that in a homogeneously expanding universe any one of its
finite parts can be regarded as a closed system to a very good
approximation, so that we can calculate the total energy and
entropy for any large enough part without having to know what lies
beyond. The relation found by this procedure will be valid for the
whole universe, at least so far as the cosmological principle can
be trusted.

With this somewhat debatable interpretation we find $\kappa =1$
for the matter-dominated era. However, in the matter-dominated era
the particles are not relativistic, so the formula for an
ultra-relativistic gas cannot be applied.

Yet when we look far back into the past where the universe is
expected to be radiation-dominated, an ideal gas of
ultra-relativistic fermions and bosons should be a very good
approximation to the thermodynamics in the very early history of
the universe. Can we attribute a sensible value to $\kappa$ in
this case?

Assuming that the expansion remains adiabatic\footnote{where
should heat and/or entropy "flow" to, when the nearly spatially
homogeneous space itself expands?}, we find that the energy
density scales with $T^4$ and the entropy-density with $T^3$. This
gives us $s \propto e^{3/4}$, which - naively - could be
interpreted as $\kappa = 4/3$. The problem is, that in the
radiation dominated era the "total energy" in any (finite)
co-moving volume element diverges ($E \propto T$), whereas the
"total entropy" in the same volume remains constant. With respect
to the total energy and entropy in a given co-moving volume
element one would rather have to postulate $\kappa \propto \ln{T}
\propto \ln{E}$, i.e. a non-constant value of $\kappa$ which
depends logarithmically on the temperature. This construct is
highly problematic. It requires a maximum temperature, because for
unbounded temperatures the ceiling $\kappa \leq 4/3$ would be
exceeded.\footnote{One could turn the argument around and argue
for a maximum temperature.} Furthermore, the entropy cannot be a
function exclusively of the total energy. This would require
$\kappa$ to be a function of total energy alone, i.e. $\kappa = c
\ln{E}$. But the "constant" of proportionality $c$ depends on the
particular choice of the co-moving volume. Unless there is a
"preferred" volume element, such as the total volume of a closed
universe, the choice of the correct co-moving volume is ambiguous.

\subsection{\label{sec:hol:univ}Holographic homogeneous flat universes}

An interesting case is that of a homogeneously expanding flat
($\Omega=1$) universe with the additional assumption, that the
matter in this universe strictly obeys the holographic principle.
This means that the entropy in any space-time region doesn't scale
with volume, but with the area of it's boundary. If we denote the
co-moving length by $r$, we find the following dependence for the
entropy in a holographic flat universe:

\begin{equation} \label{S:w}
S \propto r^2 \propto A \propto V^{\frac{2}{3}}
\end{equation}

For the total matter in a holographic universe let us assume an
equation of state of the following general form:

\begin{equation} \label{eq:EOS:hol}
P = w \frac{E}{V}
\end{equation}

In the radiation-dominated era we have $w = 1/3$, in the
matter-dominated era $w=0$. A vacuum-dominated era is
characterized by $w=-1$, a string-dominated era by $w=-1/3$ and a
domain-wall dominated era by $w=-2/3$.

Note that $w \neq 1/3$ is not necessarily in contradiction with
the equation of state for an ultra-relativistic gas. With $w$ we
denote the equation of state for {\em all} types of matter/energy
in the universe. Although matter at ultra-high temperatures will
almost certainly include a significant contribution of radiation,
the matter must not necessarily consist {\em exclusively} out of
radiation. If string theory is the correct description of the
phenomena at the high energy limit, we should expect a significant
"string contribution" to the mass-energy at ultra-high
temperatures. In this respect it is quite remarkable, that by
combining a vacuum stress-energy tensor ${T_{\mu \nu}} \propto
diag (1, -1, -1, -1)$ with the stress energy-tensor for radiation
${T_{\mu \nu}} \propto diag (1, 1/3, 1/3, 1/3)$ one gets the
stress-energy tensor of an isotropic string gas, whenever the
radiation and vacuum contributions are equal:

\begin{equation}
{T_{\mu \nu}}_{vac} + {T_{\mu \nu}}_{rad} \propto diag (1, -1/3,
-1/3, -1/3) \propto {T_{\mu \nu}}_{string}
\end{equation}

with

$$\rho_{vac} = \rho_{rad} = \frac{\rho_{string}}{2}$$

Therefore, a universe with an overall string equation of state can
still contain a significant fraction of radiation, if the
radiation is paired with an equivalent contribution of "vacuum
energy".

If one knows the equation of state for the total matter state one can
calculate the dependence of the total energy on the volume via the
thermodynamic relation $dE = - P dV$, replacing $P$ with equation
(\ref{eq:EOS:hol}):

\begin{equation}
\frac{dE}{E} = -w \frac{dV}{V}
\end{equation}

The above equation can be easily integrated, yielding

\begin{equation} \label{eq:E=V:w}
E \propto V^{-w}
\end{equation}

Relating this to equation (\ref{S:w}) we get the following
correspondence between $S$ and $E$:

\begin{equation} \label{eq:S=E:w}
S \propto E^{-\frac{2}{3 w}}
\end{equation}

so that we can identify $\kappa$ with the equation of state
parameter $w$:

\begin{equation} \label{eq:kappa:w}
\kappa \longleftrightarrow -\frac{3 w}{2}
\end{equation}

An interesting case is $w=-1/3$, which corresponds to a string
equation of state. In this case $\kappa = \frac{1}{2}$, so that

\begin{equation}
E \propto V^{\frac{1}{3}} \propto \sqrt{A} \propto r
\end{equation}

and

\begin{equation}
S \propto V^{\frac{2}{3}} \propto A \propto r^2
\end{equation}

These relations are identical to the results for a spherically
symmetric black hole, when the scale-factor $r$ is identified with
the Schwarzschild radius. We arrive at the remarkable conclusion,
that a {\em holographic} flat universe with a {\em string}
equations of state for the {\em total} matter-content delivers the
same dependence between total energy $E$, area $A$ and entropy $S$
as a spherically symmetric black hole. It is well known that the
entropy and temperature for an extreme or nearly extreme black
hole (including the correct grey-body factors) can be derived
rigourously in the context of string theory by counting string /
brane states in the low coupling limit. The construct of a
holographic flat universe suggests another non-trivial connection
between the properties of stringy matter, the holographic
principle and black hole type objects, for which the relation $S
\propto A$ holds.

In order to see how a flat holographic universe fits into the big
bang picture of a a hot, high density initial state, one has to
analyze the dependencies of energy-density and temperature on the
scale factor $r$.

From equation (\ref{eq:E=V:w}) one finds that the energy-density
scales as

\begin{equation}
e = \frac{E}{V} \propto V^{-(w+1)} \longleftrightarrow
V^{-\frac{2}{3}(\frac{3}{2}-\kappa)}
\end{equation}

The energy-density increases with decreasing scale-factor as long
as $w>-1$. The limiting case $w=-1$ is the equation of state for a
"true" cosmological constant. $w>-1$ corresponds to the condition
$\kappa < 3/2$ via equation (\ref{eq:kappa:w}). This condition is
always fulfilled for $\kappa$ in the permissible range $(0, 4/3]$.

Using equation (\ref{eq:S=E:w}) the temperature in a flat
holographic universe can be calculated:

\begin{equation}
T = \frac{dE}{dS} \propto V^{-(\frac{2}{3}+w)} \longleftrightarrow
V^{\frac{2}{3}(\kappa-1)}
\end{equation}

The temperature increases with decreasing scale factor whenever $w
> -2/3$. The limiting case $w=-2/3$ is the equation of
state of a domain wall. A flat holographic universe expanding from
a {\em hot}, high density initial state therefore requires $w >
-2/3$, or alternatively $\kappa < 1$.

Equation (\ref{eq:kappa:w}) relates $\kappa$ to the equation of
state parameter $w$. We have already seen that the implicit
equation $\kappa(u, r) = const$ only gives solutions in the range
$0 < \kappa \leq 4/3$. But this requirement was derived in the
context of microscopic statistical thermodynamics of an {\em
ultra-relativistic ideal gas}, which has an equation of state with
$w = 1/3$. With the somewhat shaky assumption, that the formalism
developed for an ultra-relativistic gas can be extended to other
forms of matter, and that the permissible range for $\kappa = S T
/ E$ is more or less independent of the matter
temperature\footnote{The black hole solutions, and even more so
the holographic solution, provide some considerable evidence for
such an assumption: The value of $\kappa$ for a black hole only
depends on the {\em dimensionless} ratios $Q^2/A$ and $J/A$, but
not on the {\em total} energy of the system $M \propto \sqrt{A}$.
$\kappa$ covers only a very limited range $0 < \kappa \leq 1/2$
(see equation (\ref{eq:kappa:bh}) in section
\ref{sec:black:holes}). However, for any non-extreme black hole
its temperature (at infinity) is proportional to $1/M$ and thus
can cover an arbitrarily large range. This demonstrates quite
clearly that the value of $\kappa$ is neither related to the size
nor the temperature of a black hole. Yet the temperature of a
system defines the relative distribution of relativistic to
non-relativistic matter! This suggests, that the value of $\kappa$
is independent from the assumption of an ultra-relativistic gas.
The value of $\kappa$ rather has to do with the relative
contributions of electro-magnetic energy ($\propto Q^2$) and
rotational energy ($\propto J$) with respect to the total energy
$M^2 \propto A$ of a self-gravitating system.

The problem in the black hole case quite clearly is, that although
the {\em global} quantities $Q$, $J$ and $A$ are well defined,
there is no way to relate these global quantities to a meaningful
notion of {\em local} rotational or electro-magnetic energy, which
is required for a microscopic statistical thermodynamic analysis.
This is quite in contrast to the holostar solution, who's local
temperature and local (singularity-free) matter state is well
defined, everywhere. Yet viewed from the exterior space-time the
holostar has nearly identical properties to a black hole. It is
subject to the exactly the same {\em exterior} constraints as a
black hole: $\kappa_{BH} = \kappa_{holo}$, as long as the exterior
quantities $Q$, $J$ and $A$ are equal.

But in the holostar space-time there exists a one to one
correspondence between the {\em exterior global} ratios $Q^2/A$
and $J/A$ to the {\em interior local} ratios of the
energy-densities: For a charged holostar the ratio of the
electro-magnetic energy-density $\rho_{em} = E^2 / (8 \pi) = Q^2 /
(8 \pi r^2 r_h^2)$ divided by the total energy-density $\rho = 1
/(8 \pi r^2)$ is constant throughout the whole interior
space-time. Furthermore, the global {\em exterior} value $Q^2/A$
is directly related to the {\em constant} local ratio of
electro-magnetic to total energy in the holostar's {\em interior}
space-time: $\rho_{em} / \rho = Q^2 / r_h^2 = 4 \pi Q^2 / A$, as
is shown in \cite{petri/charge}. But $4 \pi Q^2 / A$ is nothing
else than the dimensionless ratio in the equation for $\kappa$
(see equation (\ref{eq:kappa:bh})).

For a rotating holostar one expects that $8 \pi J / A$ will be
related to the ratio of rotational energy-density to the total
energy-density in a similar way. If this is true, the exterior
constraints $4 \pi Q^2/ A$ and $8 \pi J / A$ completely determine
the {\em constant ratio} of electro-magnetic to rotational to
total energy-density throughout the whole {\em interior} holostar
space-time. But the temperature in the interior space-time covers
the whole range from the Planck-temperature (at the holostar's
center) to the holostar's surface temperature, which is given by
$T = \hbar / (4 \pi \sqrt{r r_0})$. Therefore $\kappa$ will be
nearly independent of temperature in the whole interior
space-time.}, one can derive a restriction on the admissible
$w$-values (for the total equation of state) in a flat holographic
universe:

\begin{equation}
-\frac{8}{9} \leq w < 0
\end{equation}

It is intriguing to interpret this result such, that in a flat
homogeneous holographic universe the only equation of state that
make sense (for the total matter contribution) is $w = -1/3$ for
stringy matter and $w = -2/3$ for matter composed out of
domain-walls.\footnote{Pressureless matter with $w=0$ must be
rejected on physical grounds. Although $w$ can come as close to
zero as one likes, this implies $\kappa \rightarrow 0$. But this
requires that $u \rightarrow \infty$ for any reasonable value of
$r$, unless we assume that there are no fermions in the gas. An
infinite chemical potential per temperature for the fermions,
however, is not possible, unless one assumes that the pressureless
matter consists only out of bosons. In this case $r=\infty$ and
$\kappa=4/3$.}

The above result relied on $0 < \kappa \leq 4/3$. If one demands a
hot initial state, the requirement $\kappa < 1$ restrains $w$
somewhat further. We find $-2/3 < w < 0$, which seems to indicate
that stringy matter is the only possible form of matter at
ultra-high temperatures in a holographic flat universe. Note, that this
does not necessarily imply that all of the matter at ultra-high
temperatures must actually be strings.

\section{\label{sec:conditions:asymmetry}Necessary conditions for a thermodynamical origin of a matter anti-matter asymmetry at ultra-high temperatures}

The purpose of this section is to analyze from the most general
point of view, what conditions have to be met that microscopic
statistical thermodynamics could be a possible cause for a matter
anti-matter asymmetry in a closed system at ultra-high temperatures.

It is not the purpose of this section to give arguments whether it
is likely, unlikely, or impossible, if these conditions actually
apply to a realistic self-gravitating object. This would be a
futile task, anyway, because - as yet - we lack a universally
accepted self-consistent description for a realistic
self-gravitating object {\em containing matter}, to which
microscopic statistical thermodynamics could be
applied.\footnote{In fact, at the present time we still cannot
know for sure what properties a {\em realistic} self-gravitating
object is going to have. The important word is "realistic". The
only self gravitating object from which we {\em know} that it
exists, is the universe. Although there are other promising
candidates for self-gravitating objects, such as black holes,
there is no proof for their existence (unless one is willing to
believe that every exact and moderately simple solution to the
field equations must be realized in the real physical world).
Furthermore, despite decades of research there are yet no
plausible answers to the most fundamental questions concerning
black holes, such as the microscopic origin of the Hawking
entropy, why trapped surfaces {\em must} form in a physically
realistic space-time, how the paradox of converting a pure state
into a mixed state (by Hawking radiation/evaporation) can be
solved, what happens at the classical space-time singularities,
etc. Although a vast amount of work has been invested into the
study of black holes, not even the greatest collective effort can
guarantee that such work will bring fruit. There are other
solutions of the field equations which deserve serious
consideration.

Sadly this position is not universally shared. It would be more
appropriate, and certainly not a disgrace to science, if those who
claim to know - possibly by divine intervention? - what solution
of the field equations was selected by nature, remember the humble
statement by one of humankind's greatest minds, Sokrates: "I know,
that I don't know." These simple words are among the most
influential in the history of humanity's intellectual
achievements. They protect us from our own arrogance and guide us
to the most effective use of our limited resources: Identifying
the {\em gaps and limitations} of our knowledge. Science is a
humble profession. A too loud beating of drums by some of its
practitioners might make us miss the more important tunes. To
speak with Feynman, science is about the joy of finding things
out, but leaving the final decision to experiment and observation.

Having said this, what is the correct way to deal with such
categorical statements as "the existence of black holes and
singularities has been proven". Do the authors of such statements
know on what assumptions the proof is based? One would hope so.
Does the reader know? More often than not the assumptions are
replaced by references to great names (such as "Oppenheimer
proved", "Hawking proved") and it is left to the initiative of the
inquisitive reader to dig deep into the literature in a personal
quest for the most important ingredients to any scientific proof:
The fundamental assumptions on which the proof is based. Although
"dust-collapse", "trapped surfaces", "isotropic pressure" (even at
the string scale?) might all be viewed as reasonable assumptions,
they are not an unescapable necessity dictated by the fundamental
laws of physics. A somewhat more modest position, which carefully
analyzes on what assumptions our reasoning is built, is called
upon, if science is to remain a credible profession,
distinguishing fact from fiction and personal beliefs.

If one analyses the so called experimental or theoretical
"existence proofs" for black holes, one will quickly realize that
the picture is not as clear as the force of such statements seems
to imply: The experimental evidence for black holes is - at best -
inconclusive. Often the detection of large compact objects with
masses above the neutron star limit is already taken as proof that
black holes are real. There is not much doubt, that the active
galactic nuclei of galaxies - and even the nuclei of other
galaxies such as the milky way - contain compact objects with
masses in the range 1 million to 1 billion $M_\odot$. However, the
argument that any compact object above the neutron star limit must
be a black hole is based on the assumption, that the pressure in a
compact self gravitating object always remains isotropic and / or
non-negative. If strings are the basic building blocks of nature,
this assumption is questionable. Strings have tension (= negative
pressure) and their equation of state is naturally and necessarily
an-isotropic.

More recently the "softness" of the spectra originating from
compact black hole type objects has been taken as experimental
evidence for an event horizon, and therefore for a black hole.
However, the argument is indirect and relies on several
assumptions. The first is that any material structure that were to
replace the event horizon must have a "hard surface" producing a
"hard" spectrum (which is not observed). The argument is
convincing. But one must keep in mind that as long as one does not
{\em know} the structure of the hypothetical object, by which the
black hole and its event horizon might be replaced, it is
difficult to infer its properties. Furthermore, the spectrum is
"hard" where it is produced: at the surface of a hypothetical
compact object. But any "hard" spectrum will become "soft" for a
far away observer, if the surface redshift is high enough.
Therefore the second assumption is that the surface-redshift of
any compact self-gravitating object, which is not a black hole,
will be limited to a few $z$. This assumption is backed by a
theorem, that "proves" $z \leq 2$ for any large massive object.
Yet the proof is based on the assumption, that the pressure of a
compact self-gravitating object always remains isotropic and/or
non-negative. Once in a while the proof implicitly assumes an
analytic (or at least twice continuously differentiable) metric.
All these assumptions are not mandatory, especially if one
believes that string theory is not just a nice mathematical
endeavor, but is relevant to the real physical world. Furthermore,
the holostar and the gravastar - both exact solutions to the field
equations - have surface redshifts that exceed $z \approx 10^{20}$
(for objects above the neutron star mass limit). They also have
negative interior pressures and metrics which are only piecewise
continuously differentiable.

One also has to be quite careful with statements about the
large-scale properties of the universe. Although it is true that
WMAP \cite{WMAP/cosmologicalParameters} and the
supernova-measurements \cite{Riess, Perlmutter/Schmidt}have vastly
increased our knowledge about the structure of the universe in the
recent years, it is worth remembering that astronomy has been
rather a succession of experimental {\em surprises} than of {\em
lasting} theoretical predictions. If we are honest we must admit,
that most of the knowledge we have gained is still very much
model-dependent. While it is true that it is possible to deduce
various {\em parameters}, such as the value of the "cosmological
constant", the "fraction of cold dark matter" etc. by assuming a
particular model, such as $\Lambda$CDM, most parameters take on
quite different values, when a different model (such as
"quintessence" or the holostar model) is used. As long as the
fundamental origin of the parameters is not known, they must be
treated as what they are: best-fit numbers in a certain model. The
model-dependency and the limitations of our knowledge can be seen
quite clearly when prior-free methods are used to deduce the large
scale equation of state of the universe. With essentially the same
data some authors find a significant time-variation in the
equation of state (see for example \cite{Alam/2003, Alam/2004}),
while others point out that the data are compatible with a "true",
time-independent value of the cosmological constant
\cite{Jassal/2004, Jonsson/2004}.} The discussion of a possible
application of the results derived in this section to some
particular solutions of the field equations is referred to the
last section.

However, in the previous section several examples of simple {\em
theoretical} self-gravitating systems have been given, for which
the entropy $S$ turned out to be an exclusive function of the
total energy $E$ (and of exterior, conserved quantities, such as
angular momentum and charge). Most of these systems were
characterized by a power-law relating $S$ and $E$. Whenever this
was the case, $\kappa$ could be identified as the exponent in the
relation $E \propto S^\kappa$.

The assumption of a power-law dependence between $S$ and $E$
appears to be applicable to a wide range of self-gravitating
systems, regardless of their size. This suggests the conjecture,
that any self-gravitating system might be characterized by such a
power-law, and that different realizations of self-gravitating
systems can be classified by their specific value of $\kappa$.

Whenever $\kappa$ falls into the allowed range $0 < \kappa \leq
4/3$, one can determine the chemical potential per temperature $u$
of the ultra-relativistic fermions, when the ratio $r_f$ of
bosonic to fermionic degrees of freedom is known. The fermionic
chemical potential is generally non-zero and proportional to the
temperature, except for the rather special case $\kappa = 4/3$.

\begin{figure}[ht]
\begin{center}
\includegraphics[width=12cm, bb= 23 504 496 820]{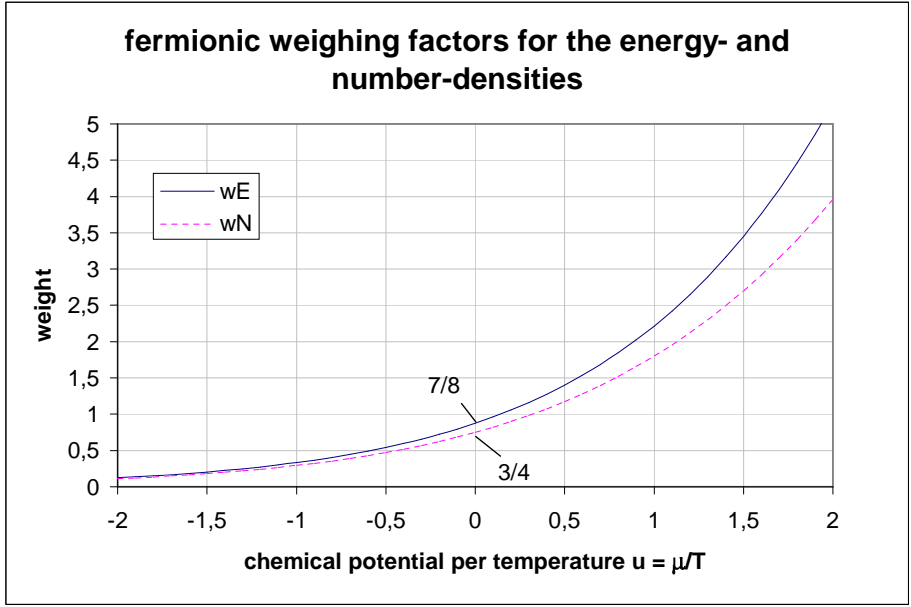}
\caption{\label{fig:asym}Fermionic weighting factors for the
energy density $w_E$ and number-density $w_N$ as a function of the
chemical potential per temperature $u$, normalized to energy- and
number-densities of a single bosonic degree of freedom}
\end{center}
\end{figure}

A non-zero value of $u$ at ultra-high temperatures is very
interesting with respect to the matter-antimatter asymmetry, which
can be found in our universe today. A non-zero chemical potential
naturally induces an asymmetry between fermions and anti-fermions.
This can be seen in Figure \ref{fig:asym}, where the number- and
energy-densities of a single fermionic degree of freedom is
plotted as a function of $u$.

As can be seen from Figure \ref{fig:asym} both the number- and
energy-densities of the fermion gas are monotonically increasing
functions of $u$. With the convention that $u>0$ for fermions and
$-u<0$ for anti-fermions one can easily see, that the number- and
energy-densities of the ultra-relativistic fermions will always be
larger than those of the anti-fermions for any non-zero value of
$u$. The extent of the asymmetry depends on $u$. The higher $u$
gets, the larger the interval $[-u, u]$ separating fermions from
anti-fermions becomes, with the result that the asymmetry
increases. Except for very small values of $u$ the asymmetry is
substantial. Let us denote the ratio of the fermionic
energy-density to that of the anti-fermions by $\eta_E$:

\begin{equation}
\eta_E = \frac{w_E}{\overline{w_E}} = \frac{w_E(u)}{w_E(-u)}
\end{equation}

When $u$ is of order unity, this ratio is already as high as
$\eta_E \approx 7.3$. The ratio of the number-densities is
somewhat lower, $\eta_N \approx 6.7$.

\begin{table}
\begin{center}
\begin{tabular} {c|c||c|c|c|c}
$\kappa$ & $f_B/f_F$ & $u$ & $\sigma$ & $\eta_E$ & $\eta_N$ \\
\hline \hline
 1.3    & 0    & 0.107512 & 4.19374 & 1.93178 & 1.87925 \\
 1.3    & 1/3  & 0.126313 & 4.01018 & 2.16376 & 2.09484 \\
 1.3    & 1    & 0.157276 & 3.85278 & 2.60340 & 2.50067 \\
 1.3    & 10   & 0.376545 & 3.63858 & 8.84216 & 8.03993 \\
 1.3    & 1000 & 2.740314 & 3.59779 & 20350.5 & 11335.5 \\ \hline
 1.2    & 0    & 0.736353 & 3.87561 & 4.05095 & 3.81702 \\
 1.2    & 1/3  & 0.856558 & 3.73527 & 5.08402 & 4.74221 \\
 1.2    & 1    & 1.047198 & 3.61357 & 7.28068 & 6.68068 \\
 1.2    & 10   & 2.160968 & 3.45176 & 57.0708 & 47.0928 \\
 1.2    & 1000 & 8.470505 & 3.48670 & 1314228 &  549864 \\ \hline
 1      & 0    & 1.344163 & 3.37921 & 12.6967 & 11.3453 \\
 1      & 1/3  & 1.531235 & 3.29933 & 17.9802 & 15.7897 \\
 1      & 1    & 1.813799 & 3.22988 & 30.2894 & 25.8847 \\
 1      & 10   & 3.289695 & 3.15745 & 420.603 & 306.846 \\
 1      & 1000 & 11.33153 & 3.33078 & $6.6 \cdot 10^7$ & $2.2 \cdot 10^7$ \\ \hline
 0.75   & 0    & 2.224800 & 2.75355 & 64.0613 & 52.5125 \\
 0.75   & 1/3  & 2.444664 & 2.73147 & 95.1571 & 76.2157 \\
 0.75   & 1    & 2.770624 & 2.71499 & 169.932 & 131.383 \\
 0.75   & 10   & 4.460321 & 2.74705 & 3007.39 & 1918.02 \\
 0.75   & 1000 & 14.14244 & 3.10305 & $2.5 \cdot 10^9$ & $6.9 \cdot 10^8$ \\ \hline
 0.5    & 0    & 3.548946 & 2.12022 & 656.161 & 464.752 \\
 0.5    & 1/3  & 3.746269 & 2.12686 & 917.248 & 635.131 \\
 0.5    & 1    & 4.055779 & 2.14011 & 1541.90 & 1030.27 \\
 0.5    & 10   & 5.843934 & 2.24645 & 27124.5 & 14804.4 \\
 0.5    & 1000 & 17.25375 & 2.78416 & $1.2 \cdot 10^{11}$ & $2.8 \cdot 10^{10}$ \\
\end{tabular}
\caption{\label{tab:parameters} Thermodynamic parameters for an
ultra-relativistic gas of fermions and bosons for various values
of $\kappa$ and $r_f = f_B/f_F$. $u$ is the chemical potential per
temperature of the fermions, $\sigma$ is the (mean) entropy per
particle, $\eta_E$ is the ratio of the energy-density of fermions
to anti-fermions, $\eta_N$ is the respective ratio of the
number-densities}
\end{center}
\end{table}

In order to supply somewhat more accurate figures for the
following discussion I have compiled some $u$-values for various
combinations of $\kappa$ and $r_f$ in Table \ref{tab:parameters} .

One can see, that for any fixed value of $\kappa$ the asymmetry is
smallest for a gas consisting completely out of fermions and
anti-fermions ($r_f=0$). The more bosonic degrees of freedom are
added to the system ($r_f$ increasing), the larger the asymmetry
between fermions and anti-fermions becomes. This reflects the fact
that $u$ increases monotonically with $r_f$ for any fixed value of
$\kappa$.

The asymmetry also depends crucially on $\kappa$. The asymmetry
becomes more profound the smaller $\kappa$ becomes. Even for
rather large values of $\kappa \simeq 1.3$, close to the maximum
value ($4/3$), the asymmetry is substantial. In order to achieve
very small asymmetries, such as the primordial asymmetry $\eta_N
\simeq 1 + 10^{-9}$ believed to have been present before
baryogenesis in the standard big bang model, one requires $\kappa$
to be extremely close to $\kappa \simeq 4/3$. Such fine-tuning is
highly improbable from the viewpoint of microscopic statistical
thermodynamics. Furthermore, a realistic self-gravitating system
with $\kappa = 4/3$ is not known to the author.\footnote{A
self-gravitating ball of photons would have $\kappa = 4/3$. But
this does not seem like a realistic system.}

For the interesting case $\kappa = 1$ (the free energy of the
system is minimized to zero) there are at least $11.35$ fermions
per anti-fermion in thermodynamic equilibrium.

From all known self-gravitating systems black hole type objects
exhibit the smallest value of $\kappa \in [0, 1/2]$ and thus the
greatest asymmetry. For $\kappa = 1/2$ and $r_f=0$ (only fermions,
no bosons) we find $\eta_N \approx 465$ and $\eta_E \approx 656$
from Table \ref{tab:parameters}. In the "supersymmetric" case
$r_f=1$ (equal numbers of bosonic and fermionic degrees of
freedom) the asymmetry is higher still, $\eta_N \simeq 635$ and
$\eta_E \simeq 917$. For black hole type objects with appreciable
angular momentum and charge $\kappa \rightarrow 0$, so that the
asymmetry becomes even more profound.

\begin{table}
\begin{center}
\begin{tabular} {c||c|c|c|c|c|c}
$u$ & $w_E$ & $\overline{w_E}$ & $w_N$ & $\overline{w_N}$ & $\sigma_F$ & $\overline{\sigma_F}$ \\
\hline \hline
 0    & 7/8     & 7/8     & 3/4     & 3/4     & 4.20183 & 4.20183 \\
 0.1  & 0.96221 & 0.79540 & 0.82138 & 0.68439 & 4.11908 & 4.28572 \\
 0.3  & 1.16213 & 0.65657 & 0.98316 & 0.56888 & 3.95720 & 4.45675 \\
 0.5  & 1.40108 & 0.54130 & 1.17339 & 0.47185 & 3.80042 & 4.63175 \\
 1    & 2.21581 & 0.33259 & 1.80039 & 0.29330 & 3.43261 & 5.08405 \\
 2    & 5.28158 & 0.12401 & 3.95683 & 0.11075 & 2.80738 & 6.03266 \\
 3    & 11.6616 & 0.04586 & 7.89005 & 0.04116 & 2.32318 & 7.01228 \\
 5    & 44.8024 & 0.00622 & 24.1791 & 0.00560 & 1.67347 & 9.00168 \\
 10   & 462.715 & 0.00004 & 152.336 & 0.00004 & 0.93967 & 14.0000 \\
\end{tabular}
\caption{\label{tab:u} Thermodynamic parameters of an
ultra-relativistic gas of fermions and bosons as a function of the
chemical potential per temperature $u$. $w_E$ is the ratio of the
energy density of the fermions, normalized to the energy density
of an ideal boson gas. $\overline{w_E}$ is the respective ratio
for the anti-fermions. $w_N$ and $\overline{w_N}$ are the
normalized number-densities. $\sigma_F$ is the entropy per fermion and
$\overline{\sigma_F}$ the entropy per anti-fermion.}
\end{center}
\end{table}

Note, that all the relevant thermodynamic parameters, such as
number-, energy- and entropy-densities of a single fermionic
degree of freedom can be calculated just by knowing $u$. In Table
\ref{tab:u} the weighting factors $w_E$ and $w_N$ and the entropy
per fermion $\sigma_F$ were compiled (barred quantities for the
anti-fermions) for different values of $u$

\section{\label{sec:critical:comparison}A critical comparison of the solutions describing a self-gravitating object}

The reader might question whether the thermodynamic model of an
ultra-relativistic gas with the additional constraint $S = S(E)$
has any relevance to the real world. It seems difficult to fulfill
both conditions under which the thermodynamic model discussed in
the previous sections is valid:

Although the condition $S = S(E) \propto E^2$ is fulfilled for a
spherically symmetric (uncharged) black hole\footnote{And is
fulfilled approximately for charged, rotating black holes.}, we
quite definitely cannot describe a black hole's interior by an
ultra-relativistic gas.\footnote{Note, however, that an observer
hovering above the event horizon will experience an intense
radiation bath due to Unruh-acceleration. The temperature of this
bath diverges, when one approaches the horizon.} The interior of a
classical black hole is vacuum (except for the "position" of the
singularity, which cannot be regarded as a well defined part of the
space-time). And whereas the assumption of an ultra-relativistic
gas appears to be valid for the early universe, the first
condition $S = S(E)$ is meaningless in the radiation dominated
phase of a homogeneously expanding FRW-type universe, as $S =
const$ in any co-moving volume, whereas $E$ diverges. The
matter-dominated phase of the universe isn't too encouraging
either: Although $S \propto E$ in any co-moving volume, the
particles are not relativistic.

However, there are situations in which both conditions might be
fulfilled. One such - rather theoretical - situation was discussed
in section \ref{sec:hol:univ}. In a flat holographic FRW-type
universe one can derive a permissible range for the equation of
state parameter $w$, by the requirement that the expansion started
out from a hot high-density initial state. We found $-2/3 < w <
0$, which points to stringy matter as the preferred form of matter
at ultra-high temperatures. An equation of state with $w=-1/3$ is
also interesting, because it correctly reproduces the relations
between entropy, energy and area of a (spherically symmetric)
black hole. As the stress-energy tensor for an isotropic
string-gas can be thought to be composed out of a radiation and a
vacuum-contribution, it is very well conceivable that a flat
holographic universe with an overall string equation of state will
contain an appreciable contribution of ultra-relativistic
particles at high densities. Therefore a holographic flat universe
can fulfill both conditions, which are $S = S(E)$ and an
appreciable contribution of radiation to the total energy budget
at high temperatures.

Another system, to which the results derived in this paper can be
applied, is the so called holographic solution. The holographic
solution, in short holostar, is an exact, static solution to the
field equations with zero cosmological constant. It has been
derived in \cite{petri/bh} and has been discussed extensively in
\cite{petri/hol, petri/thermo, petri/charge, petri/string}. In
fact the results presented in this paper are an offspring of the
thermodynamic analysis of the holographic solution reported in
\cite{petri/thermo}, applied to a broader context.

The essential difference between the holographic solution and the
- hypothetical - model of a flat holographic universe with a
string equation of state discussed in section \ref{sec:hol:univ},
is that the holographic solution describes a {\em locally anisotropic}
arrangement of strings with a well defined center and a
spherically symmetric - but inhomogeneous - energy-density,
whereas the model of a flat holographic universe is of the
isotropic, homogeneous FRW-type.

It is not possible to discuss the holographic solution in detail
in this paper. The interested reader is referred to the mentioned
citations. Here is a short summary of the results:

The holographic solution describes the most compact
self-gravitating object possible, which is {\em not} a black hole.
The holostar's properties are nearly identical to those of a black
hole as seen from the outside, but very different from the inside.
Instead of the event horizon the holostar has a real physical
boundary membrane consisting of pure tangential pressure and
situated roughly two Planck lengths outside of the holostar's
gravitational radius. The membrane separates the singularity-free
interior matter distribution from the exterior vacuum space-time.
The tangential pressure of the membrane is exactly equal to the
pressure of the - fictitious - membrane attributed to a black hole
by the membrane paradigm. This guarantees, that the dynamic
behavior of the new solution with respect to the outside world is
nearly identical to that of a black hole of the same mass.
Furthermore, the membrane is situated at the global minimum of the
gravitational potential.

The interior of the holographic solution is non-singular. This
makes it possible to describe the interior matter state by
microscopic statistical thermodynamics. A very simple
thermodynamic model discussed in \cite{petri/thermo} is the
assumption of an ultra-relativistic gas, which requires $\kappa=1$
if the thermodynamic entropy of the holostar is to reproduce the
Hawking result.\footnote{$\kappa=1$ can also be shown for a
matter-dominated holostar, if $u \ll m/T$ so that the entropy per
massive particle $\sigma \approx m/T$. See \cite{petri/hol}.} In
the holostar space-time (and in general for any thermodynamic
system with $\kappa < 4/3$) the equation $\kappa(u, r) = const$
only has a solution, if there is {\em at least one} fermionic
species present. If the gas consists only of bosons, there is no
solution. This can be interpreted such, that the degeneracy
pressure of at least one ultra-relativistic fermionic species is
required to prevent the holostar space-time to undergo continued
gravitational contraction to a singularity.

The holographic solution has a very strong string character.
Therefore it appears not justified to interpret its interior
matter state exclusively in terms of particles. If one interprets
the interior matter state in terms of strings, one finds that the
holographic solution describes a spherically symmetric, radial
collection of strings, which are attached to the spherical
boundary membrane. The strings are densely packed, in the sense
that the transverse area surrounding each string is equal to the
Planck-area. The number of strings (or rather of string segments)
attached to the spherical boundary area is equal to its proper
area, measured in Planck units. The same result applies to any
concentric sphere in the holostar's interior, meaning that the
number of strings puncturing any thin spherical concentric shell
is proportional to the proper area of the shell. In the string
picture it is very easy to see, that the number of fundamental
degrees of freedom scales with area, not with volume. Furthermore,
if one takes into account that string theory predicts a minimum
transverse area of the strings, the dense package of the strings
in the holographic solution (each separated by a Planck area from
its neighbors) is the fundamental reason, why the holographic
solution does not collapse to a singularity, although it's
boundary lies only two Planck lengths outside of it's
gravitational radius. This argument is independent of the size of
the holostar, so that arbitrarily large holostars approaching or
even exceeding the size of the observable universe can be
constructed, simply by laying out a sufficiently large number of
strings radially.

Although the overall interior equation of state is that of a
stringy matter, the holographic solution has a definite particle
interpretation: The number of ultra-relativistic interior
particles in the holographic solution can be shown to be
proportional to the area of its boundary. This result only
requires the Einstein field equations, spherical symmetry, an
interior matter-distribution $\rho = 1/(8 \pi r^2)$ and
microscopic statistical thermodynamics of an ideal
ultra-relativistic gas (see \cite{petri/thermo} for a detailed
derivation and discussion). The holographic solution therefore can
be seen as the most radical fulfilment of the holographic
principle: The number of its (interior) fundamental constituents,
be it strings or be it particles, scales with boundary
area.\footnote{For massive particles the picture is not so simple.
If one assumes that the particle masses remain constant during the
expansion, the total number of massive particles in a
matter-dominated holostar is proportional to $N_m \propto
A^{3/4}$. However, the entropy of a massive particle is equal to
$\sigma_m = m/T$, as will be shown in the Appendix. (At least this
is the case when the chemical potential per temperature $u$ is
small, which is to be expected, because the free energy for an
ideal gas of non-relativistic particles is zero for $u=1$.) With
$1/T \propto  \sqrt{r} \propto A^{1/4}$ the entropy of a
matter-dominated holostar turns out $S \propto N_m \sigma_m
\propto A$.}

Quite interestingly and unexpectedly, the holographic solution
turned out to be an excellent description for many phenomena
encountered in our universe today. Geodesic motion of particles
within the holostar space-time is practically indistinguishable
from a homogeneously expanding FRW-type universe at late times. A
geodesically moving observer will experience an isotropic
Hubble-flow of massive particles in his frame of reference. The
matter-density decreases with $\rho \propto 1/t^2$. Within the
local Hubble-volume of an geodesically moving observer the
matter-density is homogeneous by all practical purposes. The
observer is immersed in a radiation-bath, whose temperature falls
off over time with $T \propto 1/\sqrt{t}$. The geodesic motion of
photons preserves the Planck-distribution, so the radiation
remains thermal after decoupling.

Although the holographic solution has practically no tunable
parameters, it fits the observational facts remarkably well. The
overall string nature of the solution automatically leads to the
prediction of unaccelerated expansion with $H = 1 / t$, which
is in very good agreement with the observational results. Other
non-trivial relations between the Hubble length, the total
matter-density, the CMBR-temperature and the Hubble-constant can
be derived, which are all fulfilled by the observations to an
accuracy of a few percent.

\section{\label{sec:alternative:scenario}An alternative scenario for the origin of the matter-antimatter asymmetry in our universe}

The most interesting fact about the holographic solution - from the
perspective of this paper - is that it provides a promising
alternative scenario for the matter-antimatter asymmetry in our
universe. This scenario allows us to explain today's high value of
the photon-to-baryon ratio in a very natural way.

The crucial observation is, that in the holographic solution not
the number-ratios of the different particle species, but rather
their respective energy- (and entropy-) densities remain constant
during the evolution. This points to a very different type of
phase-transition at the time of nucleosynthesis and baryogenesis,
as compared to the standard cosmological models:

Today's observations show, that the energy-densities of photons
electrons are nearly equal, within a factor of 2 or 3. For any
model of the universe we can extrapolate the energy-densities back
to a time of the phase transition, which converted the
relativistic electron-positron gas to a non-relativistic electron
gas. In the standard cosmological model this extrapolation gives
the result, that the energy density of the photons must have been
a factor $10^9$ higher than that of the electrons shortly {\em
after} the phase transition. From equilibrium thermodynamics we
know that the number- and energy-densities of photons and
electrons must have been {\em nearly equal shortly before} the
phase-transition. Thus the phase-transition produced a huge
discrepancy between electron and photon numbers shortly after. The
standard explanation for this phenomenon is a very small
"primordial" asymmetry of the order $1 + 10^{-9}$ between
electrons and anti-electrons. However, this is an "a posteriori"
explanation. So far we lack a theory which would allow us to
predict the correct theoretical value of the asymmetry from first
principles.

If we extrapolate today's observed energy-densities back to the
time of the phase-transition using the holostar model, we find
quite a different result: The energy-density of photons and
electrons are comparable, {\em before and after} the phase
transition.  This means, that instead of producing a great
asymmetry in the number- and energy-densities of photons and
electrons, the positrons dropped out rather smoothly during the
phase-transition. It is quite clear, that such a smooth
transition, which leaves the energy- and number-densities of the
different particle species nearly unchanged, requires a rather
{\em large} asymmetry between fermions and anti-fermions shortly
before the phase-transition. However, this is exactly what the
thermodynamic analysis of the holographic solution
predicts.\footnote{More generally: Any thermodynamic closed system
at high temperatures with $E \propto S^{\kappa}$ and $\kappa$ well
below $4/3$ predicts such a large asymmetry.}

For the following discussion let us assume $\kappa=1$, which is
required if the holostar-solution is to reproduce the Hawking
temperature and entropy correctly (or if the free energy of the
system should be minimized to zero).

What will the equilibrium ratio of electrons to positrons be at
the electron-mass threshold $T \approx m_e/3$ for $\kappa=1$? To
answer this question we have to determine $u$, for which we have
to know the ratio $r = f_B/f_F$ of bosonic to fermionic degrees of
freedom. Slightly above the electron-mass threshold the only
relativistic particles will be electrons, positrons, neutrinos and
photons.  The two helicity-states of the electron and the three
neutrino-flavors (each with one helicity state) amount to $f_F =
5$. The photons are characterized by $f_B=1$. This results in $r =
1/5$ for which we find $u \approx 1.46$.

The ratio of electron and positron number-densities only depend on
$u$ and are given by

\begin{equation}
\eta_N = \frac{Z_{F,2}(u)}{Z_{F,2}(-u)} \approx 14
\end{equation}

There will be roughly 14 electrons per positron in thermal
equilibrium. When the temperature falls below the threshold, one
electron annihilates with one positron, leaving roughly 13
electrons behind. The annihilation energy, which is roughly $2/15
\approx 13 \%$ of the original energy, will be distributed among
the remaining ultra-relativistic particles, which are photons and
neutrinos. If the neutrinos have already decoupled from the
radiation, all of the energy will go into the photons.

In thermodynamic equilibrium well above the threshold the
energy-density of the electrons and anti-electrons is roughly
twice as high as the energy-density of the photons for $u \approx
1.46$:

\begin{equation}
\frac{e_{e^\pm}}{e_\gamma} = \frac{w_E +
\overline{w_E}}{2} \approx 1.77
\end{equation}

The annihilation diminishes the energy-density in the electrons by
a fraction $2 \overline{w_E} / (w_E + \overline{w_E}) \approx 12
\%$.

For simplicity let us assume that all of the annihilation energy
goes to the photons. This would be the case if the neutrinos had
already decoupled from the radiation. As the energy density of the
photons is roughly half the energy-density in the electrons, the
transfer of roughly 12 \% of the electron energy-density to the
photon gas will increase the energy-density of the photons by
roughly 20 \%. The ratio of the energy-densities of electrons and
photons after the annihilation-process will be roughly given by

\begin{equation} \label{eq:energy:ratio}
\frac{e_{e}}{e_\gamma} = \frac{w_E -
\overline{w_E}}{2 + 2 \overline{w_E}} \approx 1.29
\end{equation}

We find, that the energy-densities of electrons and photons are
still comparable to each other after annihilation. The
phase-transition proceeds quite smoothly in a quasi-equilibrium
way.

The numerical figures quoted here are just rough estimates.
However, due to the quasi-equilibrium nature of the phase
transition one can be quite confident, that equilibrium
thermodynamics is a fairly good approximation throughout the whole
process. Therefore the found ratio $e_e / e_\gamma \approx 1.3$
should not deviate greatly from the exact result. A detailed
calculation is expected to give corrections which might raise or
lower the estimate by a factor of $2$, but not vastly more.

If the ratio of the energy-densities is conserved in the further
evolution of the universe, as is the case for particles in
geodesic motion within the holostar-solution, one gets the
prediction, that the electron-density should be roughly a factor
$1.3$ higher than the photon density today. The actual value
appears to be somewhat lower: If we take the WMAP values $\Omega_b
\approx 0.04$ and $h = 0.71$ and a proton to nucleon-ratio of 7/8
we get

\begin{equation}
\frac{e_{e}}{e_\gamma} \approx 0.4
\end{equation}

This is a factor of 3 too low. On the other hand, the value of
$\Omega_b$ has been determined in the context of an FRW-model. Any
such determination is model-dependent, as can be seen quite
clearly for the case of the so-called "dark energy". In order to
explain the observations in the standard FRW-models we require a
significant dark-energy component $\Omega_\Lambda \approx 0.7$,
which is usually identified with a positive cosmological constant.
In the holostar-model the same phenomena are explained in the
context of an exact solution of the field equations with {\em
zero} cosmological constant. At first order the
positive cosmological constant in the FRW-model has the same
effect on the luminosity-redshift relation as the string equation
of state in the holostar-model.\footnote{At second order there are
differences. However, with today's experimental data it is
difficult to decide between two models on the basis of second
order effects.}

It is conceivable, that the holostar model of the universe doesn't
need dark matter either, or maybe not as much. In such a case the
electron density could be higher, up to a factor of 6. Today's
observational value for the ratio of electron to photon
energy-density then would lie in the range 0.4 \ldots 2.4.

In the holostar solution the ratio of the energy-densities remains
constant. This allows us to predict the baryon to photon ratio
from first principles. As the individual photon energy scales with
temperature, the number-density of the photons must increase (with
respect to the electrons) when the temperature is lowered. If one
knows the ratio of the energy- and number densities at any one
time, one can calculate the baryon-to-photon ratio as a function
of $T$. With $n_b \approx (8/7) \, n_e = (8/7) e_e / m_e$ and the
well known relation between energy-density and number-density of a
photon gas $e_\gamma \simeq 2.7 \, n_\gamma T$ we find:

\begin{equation}
\frac{n_b}{n_\gamma} \approx \frac{3 T}{m_e}
\frac{e_{e}}{e_\gamma}
\end{equation}

If we replace $T$ with the current value of the CMBR-temperature
and plug in the prediction from equation (\ref{eq:energy:ratio})
for the ratio of the energy-densities we can estimate the
baryon-to-photon ratio today. It turns out as:

\begin{equation}
\eta \approx 18.3 \cdot 10^{-10}
\end{equation}

which is a factor of 3 higher than the baryon-to-photon ratio
determined by WMAP.

\section{Discussion}

The very simple model of an ultra-relativistic gas described by
three dimensionless parameters, the number of fermionic degrees of
freedom $f_F$, the number of bosonic degrees of freedom $f_B$ and
the chemical potential per temperature of the fermions $u$
naturally leads to a profound matter-antimatter asymmetry,
whenever $u$ is non-zero.

For self-gravitating systems characterized by a power-law
dependence between total energy $E$ and entropy $S$, i.e. $E
\propto S^\kappa$ the parameter $u$ can be determined when the
ratio $r_f = f_B / f_F$ of bosonic to fermionic degrees of freedom
is known. $\kappa = S T / E$ has been shown to be a symmetric
function of $u$ at ultra-high temperatures. Whenever the implicit
equation $\kappa(u, r_f) = const$ has a non-zero solution, we find
two values for $u$, one positive, the other equal in magnitude but
negative. Positive $u$ describes fermions, its negative
counterpart anti-fermions. Solutions only exist in the range $0 <
\kappa \leq 4/3$. The maximum possible value for $\kappa$ is
$4/3$, in which case $u$ is zero regardless of $r_f$. This
corresponds to the well known case of an ultra-relativistic gas of
fermions and bosons with zero chemical potential. For $\kappa <
4/3$ the value of $u$ is always non-zero, and grows with
increasing $r_f$.

A non-zero value of $u$ leads to a natural asymmetry between
fermions and anti-fermions in thermodynamic equilibrium. This
asymmetry is profound unless $u$ is very small (which requires
$\kappa \rightarrow 4/3$).

Several examples for self-gravitating systems were given, which
have values of $\kappa$ in the range between $0 < \kappa \leq 1$.
A spherically symmetric Schwarzschild black hole corresponds to
$\kappa = 1/2$. Charged and rotating black holes have lower
values. Any extreme Kerr-Newman black hole has $\kappa \rightarrow
0$. The black holes exhibits the largest asymmetry among the known
types of self-gravitating objects.

By the definition of $\kappa$ the Gibb's free energy $F$ is
related to the total energy by $F = (1-\kappa) E$. Therefore,
minimizing the free energy of a system with a given total energy
$E$ corresponds to maximizing $\kappa$. In the black hole case,
the free energy is minimized for a Schwarzschild black hole: $F
=E/2$ with $\kappa = 1/2$. However, there are self-gravitating
objects which minimize the free energy even further.

The recently discovered holographic solution is characterized by
$\kappa = 1$ in its matter-filled (interior) region. These
solution has an interior matter state which is non-singular, so
that the assumption of an ultra-relativistic gas is valid, at
least for small values of the scale factor and high densities. For
$\kappa =1$ we have an asymmetry at high temperatures which
amounts to 1 anti-fermion per 14 fermions at a temperature
corresponding to the electron-mass threshold. When the temperature
drops below this threshold, the mutual annihilation of electrons
and positrons is very moderate. 1 positron annihilates with 1
electron leaving 13 electrons behind. Assuming that the
annihilation proceeds in a quasi-equilibrium way, the equilibrium
ratio of the energy-densities of electrons to photons before
annihilation is roughly $1.8$, after the annihilation roughly 1.3.

In the holographic solution it is not the number-densities of the
particles that remain constant during the expansion, but the
energy- and entropy-densities. This means, that the number-density
of the electrons with respect to the number-density of the photons
falls linearly with temperature. From this behavior the
baryon-to-photon ratio at the CMBR-temperature can be predicted.
It amounts to $\eta \approx 18 \cdot 10^{-10}$, quite close to the
current value of $\eta \approx 6.5 \cdot 10^{-10}$.

The simple thermodynamic model of an ultra-relativistic gas
combined with the holographic solution allows us to construct a
very different scenario of baryogenesis / nucleosynthesis, which
proceeds from a profound - thermodynamically induced -
matter-antimatter asymmetry to the nearly complete absence of
anti-matter today. Because the asymmetry was profound already at
high temperatures, the phase-transitions at the respective
mass-thresholds of the particles proceed very smoothly, as the
fraction of anti-matter that is annihilated at the threshold is
quite small (less than 8\% of the total). This scenario has some
experimental backing in today's observed baryon to photon number
ratio.

A straight forward task for future research is a more accurate
calculation for the energy- and number-densities of the electrons
and photons after the annihilation of the positrons at the
electron mass threshold, taking into account all relevant
interactions.

A more difficult task will be the analysis, how nucleosynthesis
proceeds in the holostar universe. Here one has to take two
effects into account. First, the Hubble-rate at the
nucleosynthesis temperature in the holostar universe is
significantly higher than in the standard FRW-models. A higher
expansion rate shuts off the reactions converting D, T and He3 to
the tightly bound He4 nucleus faster, so that a higher fraction of
D and He3 will be left over. On the other hand, the baryon and
electron number densities at the temperature of nucleosynthesis is
higher than in the FRW-models, which makes the conversion to He4
more effective. The higher density counteracts the faster
expansion rate, so it is conceivable that nucleosynthesis in the
holostar universe might produce results similar to the
calculations based on the standard FRW-model. When we have a good
picture how nucleosynthesis proceeds in the holostar model of the
universe, we might be able to decide whether the scenario for the
matter-antimatter asymmetry presented in this work is worthwhile
of further consideration, not only from a theoretical perspective,
but from a very practical point of view.


\newpage
\appendix

\section{Thermodynamics of an ideal gas}

In this appendix I summarize the basic derivations and results for
the thermodynamics of an ideal gas of non-interacting particles,
which are subject to the exact relativistic energy-momentum
equation $\epsilon^2 = p^2 + m^2$. The momentum distribution is
assumed to be spherically symmetric. The treatment is fully
general and complete. It includes the possibility of a non-zero
chemical potential of the particles. The derivations are exact.
The results involve integrals which cannot be expressed in terms
of simple functions. For the two relevant limiting cases, the
ultra-relativistic case $p \gg m$ and the non-relativistic case $p
\ll m$, the results can be approximated to an excellent precision
by standard mathematical functions, such as the Gamma- or the
Poly-logarithmic function.

I haven't found this elsewhere, so this compilation might turn out
useful for future reference.

\subsection{The grand canonical potential $J$ and some useful abbreviations}

The thermodynamic relations for an ideal gas will be derived in
the grand canonical formalism. The grand canonical potential for a
single particle species with $f$ internal degrees of freedom is
given by:

\begin{equation} \label{eq:App:J}
J = \mp f \, T \int{\frac{d^3p \, d^3x}{(2 \pi \hbar)^3} \ln{(1
\pm e^{-\frac{\epsilon - \mu}{T}})}}
\end{equation}

The $+$ sign in the logarithm refers to fermions, the $-$ sign to
bosons. The sign in front of the integral is reversed, i.e. a $-$ sign
for fermions, a $+$ sign for bosons.

For an ideal non-interacting gas, which is not subject to any
exterior forces, the energy $\epsilon$ of a particle will not
depend on position $x$. The integral over $d^3x$ just gives the
spatial volume occupied by the gas:

$$V = \int{d^3x}$$

We will choose the volume small enough, so that the relevant
thermodynamic parameters $T$ and $\mu$ are effectively constant in
this volume. (For any realistic model one will have to check,
whether the solution is compatible with this assumption).

Under the assumption of a spherically symmetric momentum
distribution $\epsilon = \epsilon(p)$, with $p^2 = p_x^2 + p_y^2 +
p_z^2$, we can rewrite equation (\ref{eq:App:J}) as follows:

\begin{equation} \label{eq:App:j}
j = \frac{J}{V} = \mp \frac{f}{2 \pi^2 \hbar^3} T
\int_0^{\infty}{p^2 \ln{(1 \pm e^{-\frac{\epsilon - \mu}{T}})} dp}
\end{equation}

$j$ is the density of the grand canonical potential. The total
(extrinsic) thermodynamic quantities for a given volume $V$, such
as the grand canonical potential $J$ or the total energy $E$ will
be written with capital letters, whereas the densities will be
denoted by lower case letters. Quantities referring to the
individual particles, such as the energy per particle $\epsilon$
will be denoted by (lower case) greek letters. In the following
paragraphs the derivations will be done for a pure fermion gas.
The calculation for bosons is practically identical. One just has
to replace the $+$ sign with a $-$ sign in all results. In order
to save space I will use the abbreviation

\begin{equation}
g = \frac{f}{2 \pi^2 \hbar^3}
\end{equation}

which includes the number of degrees of freedom $f$ and the
dimensional factor $1 / (2 \pi^2 \hbar^3)$, which occurs commonly
for a spherically symmetric momentum distribution in three spatial
dimensions.

The density of the grand canonical potential for fermions $j_F = J
/ V$ then becomes:

\begin{equation} \label{eq:App:jF}
j_F = -g_F T \int_0^{\infty}{p^2 \ln{(1 + e^{-\frac{\epsilon(p) -
\mu}{T}})} dp}
\end{equation}

The energy-momentum relation for any free particle is given by:

\begin{equation}
\epsilon^2 = p^2 + m^2
\end{equation}

For the integrals that arise it is useful to replace the
dimensional integration variable $p$ with a dimensionless variable
$z$. Let us define $z$ as the "kinetic energy" (total energy minus
energy at rest) of the particle, scaled to the local temperature:

\begin{equation} \label{eq:App:z}
z = \frac{\epsilon - m}{T}
\end{equation}

It is easy to show that

\begin{equation} \label{eq:App:p2dp:1}
p^2 dp = T^3 (z+\frac{m}{T}) \sqrt{z} \sqrt{z + 2 \frac{m}{T}} dz
\end{equation}

Let us denote the $z$-dependent term in the above equation by:

\begin{equation} \label{eq:App:I'}
I'(z, x) =  (z + x) \sqrt{z (z + 2 x)}
\end{equation}

In the following discussion the dimensionless ratio $x = m/T$ is a
very important quantity:

\begin{equation} \label{eq:App:x}
x = \frac{m}{T}
\end{equation}

$x \rightarrow 0$ in the ultra-relativistic case $m \ll T$ with $x=0$
for relativistic particles, such as photons, whereas $x \gg 1$ in
the non-relativistic case.

With the above abbreviations we find

\begin{equation} \label{eq:App:p2dp}
p^2 dp = T^3 I' dz
\end{equation}

The integral of $I'$ with respect to $z$ is easy to derive:

\begin{equation} \label{eq:App:I}
I(z, x) =  \frac{\left(z (z + 2x)\right)^{\frac{3}{2}}}{3}
\end{equation}

The integration constant has been chosen such, that $I(0, x) = 0$.
This choice is dictated by the requirement, that the integral for
$j$ can be transformed to a standard form by an integration by
parts with zero boundary term.

For the following calculations the dimensionless chemical
potential per temperature $u$ will turn out useful

\begin{equation} \label{eq:App:u}
u = \frac{\mu}{T}
\end{equation}

With these abbreviations the dimensionless expression in the
exponent of the Boltzmann-factor becomes:

\begin{equation} \label{eq:App:alpha}
\alpha = \frac{\epsilon - \mu}{T} = z - \frac{\mu}{T} +
\frac{m}{T} = z - u + x
\end{equation}

Note that $\alpha$ depends linearly on the dimensionless
integration variable $z$. We are now ready to transform the
expression for the grand canonical potential by an integration by
parts to a more familiar expression:

\begin{equation} \label{eq:App:jF:2}
j_F = -g_F T^4 \int_0^\infty{I'(z,x) \ln{(1 + e^{-\alpha(z,x,u)})}
dz} = -g_F T^4 \int_0^\infty{n_F \, I dz}
\end{equation}

where $n_F$ is the so called occupation number for a fermion:

\begin{equation}
n_F = \frac{1}{1 + e^{\frac{\epsilon - \mu}{T}}}= \frac{1}{1 +
e^{\alpha}} = \frac{1}{1 + e^{z-u+x}}
\end{equation}

\subsection{Deriving thermodynamic quantities from the grand canonical potential }

The entropy-density $s = S/V$ in the grand-canonical formalism is
derived by a partial differentiation of the grand canonical
potential density by $T$:

\begin{equation}
s_F = - \frac{\partial j_F}{\partial T} = g_F \,
\frac{\partial}{\partial T}\left(  T \ln{(1+e^{-\frac{\epsilon -
\mu}{T}})} \right)
\end{equation}

from which we find (by pulling the derivative under the integral):

\begin{equation} \label{eq:App:s}
s_F = \frac{-j_F}{T} + g_F \int_0^\infty{ \frac{\epsilon - \mu}{T}
\frac{p^2 dp}{1+e^{\frac{\epsilon - \mu}{T} }}} = g_F T^3
\int_0^\infty{n_F \left(I + \alpha I' \right) dz}
\end{equation}

The number-density $n = N/V$ follows from a partial derivative
with respect to $\mu$:

\begin{equation}
n_F = - \frac{\partial j_F}{\partial \mu} = g_F T \,
\frac{\partial}{\partial \mu}\left( \ln{1+e^{-\frac{\epsilon -
\mu}{T}}} \right)
\end{equation}

which gives:

\begin{equation} \label{eq:App:n}
n_F = g_F \int_0^\infty{\frac{p^2 dp}{1+e^{\frac{\epsilon -
\mu}{T} }} } = g_F T^3 \int_0^\infty{n_F I' dz}
\end{equation}

The energy-density $e = E/V$ is given by:

\begin{equation}
e_F = j_F +  s_F \, T + n_F \, \mu
\end{equation}

which evaluates to:

\begin{equation} \label{eq:App:e}
e_F = g_F T^4 \int_0^\infty{n_F \left( x + z\right) I' dz}
\end{equation}

The pressure is given by a partial derivative with respect to the
volume:

\begin{equation} \label{eq:App:P}
P = - \frac{\partial J_F}{\partial V} = -j_F = g_F T^4
\int_0^\infty{n_F \, I \, dz}
\end{equation}

For a boson gas we can use the same expressions. We just have to
replace the fermion occupation number $n_F$ by the occupation
number for a boson:

$$n_F \rightarrow n_B = \frac{1}{e^{\frac{\epsilon - \mu}{T}}-1} = \frac{1}{e^{\alpha}-1} $$

So far the derivation was completely general. For any sufficiently
small volume, where the temperature and the chemical potential can
be considered to be constant, the relevant thermodynamic
parameters, such as entropy-, energy- and number-density, pressure
etc. are related via integrals of the form:

$$ \int_0^\infty{n_P(z, u, x) \, z^n \, I(z, x) \, dz}$$

or

$$ \int_0^\infty{n_P(z, u, x) \, z^n \, I'(z, x) \, dz}$$

These integrals only depend on the two dimensionless parameters $u
= \mu / T$ and $x = m / T$, which take on (nearly) constant values
in any sufficiently small volume, where the temperature can be
considered to be constant. $n_P$ is the occupation number, with
$P$ denoting the particle species ($n_F$ for a fermion, $n_B$ for
a boson).

For fermions the above integrals are well defined for all values
of $x$ and $u$. For bosons this is not the case, because $n_B(z)$
has a pole along the positive $z$-axis, whenever $x<u$. Therefore,
for bosons we have to postulate that $x_B \geq u_B$, which means
that the rest mass of a boson can never be less than its chemical
potential in the ideal gas case:

\begin{equation}
m_B \geq \mu_B
\end{equation}

For a gas of ultra-relativistic bosons (with $x_B=0$) we get the
already known result $u_B \leq 0$.

Many qualitative features and some quantitative results can be
obtained without having to solve the integrals. To give an
example: From equation (\ref{eq:App:P}) we find that the partial
pressure depends on the fourth power of the temperature times a
function $f$ of the dimensionless parameters $u$ and $x$:

$$P \propto T^4 f(u, x) $$

with

$$f(u,x) = \int_0^\infty{n_P \, I \, dz}$$

For a gas with zero chemical potential and for ultra-relativistic
energies it is easy to see, that $f \rightarrow const$, so we
recover the well known result $P \propto T^4 \propto e$ in the
ultra-relativistic case.

More specifically, the pressure is related to the energy-density
$e$ by:

\begin{equation}
P = w(x, u) \, e
\end{equation}

with

\begin{equation} \label{eq:App:EOS}
w(x, u) = \frac{\int{n_P I dz}}{\int{n_P (x + z) I' dz}}
\end{equation}

It is not difficult to evaluate the above integral numerically for
any given ratio of $u = \mu /T$ or $x = m/T$. For $x=0$, i.e. for
the ultra-relativistic case, $w = 1/3$ independent of $u$.

More often than not we don't require the general result. Usually
it suffices to know the (quite common) special cases of an ideal
gas at ultra-relativistic energies ($x \rightarrow 0$) or at
non-relativistic energies $x \gg 1$. For these cases it is easy to
approximate the expressions in the integrals and solve the
integrals exactly by the poly-logarithmic function and/or the
Gamma-function. This will be shown in the following section.

\subsection{\label{app:rel}The ultra-relativistic case}

For an ultra-relativistic gas the influence of the rest-mass $m$
of the particles becomes negligible, so we can set $m=0$. The
integrals only depend on one parameter, the chemical potential per
temperature $u$. We find:

\begin{equation}
I \rightarrow \frac{z^3}{3}
\end{equation}

\begin{equation}
I' \rightarrow z^2
\end{equation}

\begin{equation}
\alpha = \frac{\epsilon - \mu}{T} \rightarrow z-u
\end{equation}

\begin{equation}
n_P \rightarrow \frac{1}{e^{z-u} \pm 1}
\end{equation}

With these approximations (which are exact for zero-rest mass
particles) all integrals are of the form:

$$\int_0^\infty{\frac{z^n \, dz}{e^{z-u} \pm 1}}$$

Any of these integrals reduces to finding the respective value of
the poly-logarithmic function:

\begin{equation}
\int_0^\infty{\frac{z^n \, dz}{e^{z-u} \pm 1}} = \mp \Gamma(n+1)
PolyLog_{n+1}(\mp e^u)
\end{equation}

The minus sign is for fermions, the plus sign for bosons.

In some cases not even that is needed. For example, using equation
(\ref{eq:App:EOS}) in the ultra-relativistic case yields:

$$w \rightarrow \frac{\int{n_P \frac{z^3}{3} dz}}{\int{n_P z^3 dz}} = \frac{1}{3}$$

Not quite unexpectedly we get the equation of state for an
ultra-relativistic gas: $P = w \, e$ with $w = 1/3$. We have
already seen from the numerical evaluation of $w$ that the
equation of state is independent of the value of $u$ and
independent of the particle species (fermions, bosons) at
ultra-relativistic energies. The equation of state for an
ultra-relativistic gas is universal. This must be so: If we take
the stress-energy tensor of a homogeneous matter distribution of
massive particles $T_{\mu \nu} = diag(\rho, 0, 0, 0)$ and boost it
in the six spatial directions ($\pm x, \pm y \pm z$) with a high
$\gamma$-factor, the stress-energy tensor that results from the
addition of the six boosted versions is nothing else than the
stress-energy tensor of an ultra-relativistic gas. The same result
is obtained, when one sums up the stress energy tensor for a
homogeneous flow of photons (or neutrinos) moving in the six
possible spatial directions. From this argument one can see very
clearly, that the chemical potential of an ultra-relativistic
particle is irrelevant, so far as the equation of state is
concerned.

\subsection{\label{app:nonrel}The non-relativistic case}

In the non-relativistic case we can replace the relevant
quantities in the integrals with the following expressions:

\begin{equation}
I \rightarrow \sqrt{2} \frac{2}{3}
\left(\frac{m}{T}\right)^{\frac{3}{2}} z^{\frac{3}{2}}
\end{equation}

\begin{equation}
I' \rightarrow \sqrt{2} \left(\frac{m}{T}\right)^{\frac{3}{2}}
z^{\frac{1}{2}}
\end{equation}

\begin{equation}
\alpha = z-u+x
\end{equation}

\begin{equation}
n_P \rightarrow e^{-z} \, e^{-(\frac{m}{T}-u)}
\end{equation}

This means, that all integrals are of the form, which can be
evaluated by the Gamma-function:

\begin{equation}
\int_0^\infty{z^n \, e^{-z} \, dz} = \Gamma(n+1)
\end{equation}

When ratios of the above integrals are required, we can use the
well known relation for the Gamma-function in order to find exact
results.

$$\Gamma(n+1) = n \Gamma(n)$$

Note that in the non-relativistic case $n_P$ is independent of the
particle species. The only relevant parameters are the particle's
mass per temperature $x = m/T$ and its chemical potential per
temperature $u = \mu/T$.\footnote{In the ultra-relativistic case
the chemical potential per temperature $u$ is a good indicator for
the nature of the fundamental particle. $u=0$ for bosons, whereas
$u \neq 0$ for fermions. The argument leading to $u_B=0$ relied on
the observation, that ultra-relativistic bosons (with
$x=\rightarrow 0$) cannot have a positive chemical potential,
because the respective integrals give imaginary results whenever
$u_B>0$. As the chemical potential of particle and anti-particle
must be opposite ($u_B + \overline{u_B}=0$), this only leaves $u_B
= 0$. However, for non-relativistic energies the bosons {\em can}
have a positive chemical potential. Whenever $x>0$ the relevant
integrals give sensible results for a small range of positive
values $0 \leq u_B < u_{max}$. (Note however, that although there
is no restriction for negative values of $u_B$, there is a maximum
value on the positive side, which depends on the ratio $x = m/T$.
For $x=0$ we have $u_{max}=0$. When the bosons become
non-relativistic, i.e. $x$ increases, $u_{max}$ increases in
unison.). As the thermodynamic properties of bosons and fermions
become nearly identical for energies well below a particle's rest
mass (the $\pm1$ contribution in the Boltzmann-factor is utterly
negligible with respect to $e^{m/T-u}$), it is doubtful whether
$u$ remains a good "indicator" for the nature of the particles in
the extreme non-relativistic case.}

\subsection{Some important relations}

In the following paragraphs some important relations, such as the
entropy per particle, the (free) energy per particle etc. are
compiled. Whenever appropriate I have omitted the index $F$ or $B$
denoting a fermion or boson.

\subsubsection{\label{app:sigma}The entropy per particle}

The entropy per particle $\sigma$ is obtained by dividing the
entropy density $s$ by the number-density $n$. We find:

\begin{equation}
\sigma = \frac{s}{n} = \frac{m}{T} - u + \eta_\sigma(u, x)
\end{equation}

with

\begin{equation}
\eta_\sigma(u, x) = \frac{\int{n_P (I + zI') dz} }{\int{n_P I' dz} }
\end{equation}

Let us discuss the two relevant limiting cases in somewhat greater
detail. In the non-relativistic case it is easy to see with the
relations given in section \ref{app:nonrel}, that $\eta_\sigma
\rightarrow 5/2$, independent of $u$:

\begin{equation}
\eta_\sigma(u, x \rightarrow \infty) \rightarrow \frac{\int{e^{-z}
(\frac{2}{3} +1)z^{3/2} dz} }{\int{e^{-z} z^{1/2} dz} } =
\frac{5}{3} \frac{\Gamma(5/2)}{\Gamma(3/2)} = \frac{5}{2}
\end{equation}

Therefore the entropy per particle for an ideal gas of massive
particles at non-relativistic energies $\sigma_m$ is given by:

\begin{equation} \label{eq:App:sigma:m}
\sigma_m = \frac{m}{T} - u + \frac{5}{2}
\end{equation}

For $u \approx 0$ and for $m/T \gg 1$ we get the known result,
that the entropy per massive particle is (almost) equal to the
ratio of rest mass to temperature.

$$\sigma_m \approx \frac{m}{T}$$

The above result can be expressed somewhat differently:

\begin{equation} \label{eq:App:sigma:m:2}
\sigma_m = \frac{m + \frac{3}{2} T - \mu}{T}  + 1 = \frac{m +
\epsilon_{th} - \mu}{T}  + 1 = \frac{\epsilon - \mu}{T} + 1
\end{equation}

$\epsilon_{th} = \frac{3}{2} \, T$ is nothing else than the (mean)
thermal energy of the particle in three spatial dimensions
($\frac{1}{2} \, k T$ for each spatial dimension according to the
equipartition theorem\footnote{The mean thermal energy per
particle $\epsilon_{th} = \frac{3}{2} T$ can be derived rigorously
via $\epsilon_{th} = \int{p^2/(2m) \, n_P \, p^2 dp} \, / \,
\int{n_P p^2 dp}$.}) and $\epsilon = m + \epsilon_{th}$ is the
(mean) total energy of the particle. We get the interesting
result, that the entropy per particle is exactly one, if the
chemical potential of the particle is equal to its total energy,
i.e. $\mu = \epsilon$. On the other hand, if the chemical
potential per temperature $u$ is small, the entropy per particle
is dominated by the term $m/T$, which becomes very large for
temperatures well below the rest mass of the particle.

In the ultra-relativistic case we find:

\begin{equation} \label{eq:App:fsigma:u}
\eta_\sigma(u, x) \rightarrow \frac{4}{3} \frac{
\int{\frac{z^3}{e^{z-u} \pm 1} dz} } { \int{\frac{z^2}{e^{z-u} \pm
1} dz} }
\end{equation}

The only relevant parameter in the ultra-relativistic case is the
dimensionless quantity $u = \mu / T$. It is convenient to define
the following functions of $u$:

\begin{equation}
Z_{F,n}(u) = \int_0^\infty{z^n n_F(z,u) dz} =
\int_0^\infty{\frac{z^n\, dz}{e^{z-u}+1}}
\end{equation}

and

\begin{equation}
Z_{B,n}(u) = \int_0^\infty{z^n n_B(z,u) dz} =
\int_0^\infty{\frac{z^n \, dz}{e^{z-u}-1}}
\end{equation}

In the following discussion I will omit the index $B, F$, whenever
$Z_{F,n}(u)$ or $Z_{B,n}(u)$ is referenced. With the above
definitions we have:

\begin{equation} \label{eq:App:sigma:u}
\sigma(u) \rightarrow \frac{4}{3} \frac{Z_{3}(u)}{Z_{2}(u)} - u
\end{equation}

demonstrating that the entropy per particle in the
ultra-relativistic case depends only the dimensionless quantity $u$.

\subsubsection{The entropy per particle in general relativity}

We can express the entropy per particle in equation
(\ref{eq:App:sigma:m}) in a form which will prove most interesting
from a general relativistic viewpoint. In general relativity all
energies are treated on the same footing. The space-time metric
only depends on the total energy. There is no need to distinguish
between the "kinetic energy", the "rest mass energy" or the -
often not even definable notion of - "gravitational
energy" for a self gravitating system. This property of general
relativity is very welcome, because it allows us to calculate the
metric without having have to know, how a large (or small !)
system works internally, i.e. how the different "types" of energy
add up.\footnote{The total energy is the only relevant observable
in GR. Splitting up the total energy into different contributions
is just a convenient bookkeeping device, but shouldn't be
considered to have any deeper meaning beyond it's main purpose:
Allowing us to do gain some intuitive understanding and - in some
instances - allowing us to set up some equations. Note, that even
the apparently well defined concept of "kinetic" energy is
somewhat ambiguous from a relativistic point of view: The kinetic
energy depends on the frame of reference. The kinetic energy is
not invariant under a Lorentz boost!}

On the other hand, it is well known that self-gravitating systems,
such as black holes, have an entropy which is directly related to
the system's total energy (in the case of a spherically symmetric
black hole, its entropy is related to the square of the system's
total gravitational mass). This observation (as well as the
principle of equivalence and the scale invariance of the theory)
suggests, that the entropy of any self gravitating system,
including particles, should {\em only} depend on (i) it's total
energy $ \epsilon$ and (ii) the total energy's {\em conjugate}
thermodynamic parameter, which is temperature. The temperature is
required, because entropy is a dimensionless quantity. The only
dimensionless parameter that can be formed from the total energy $
\epsilon$ and the temperature $T$ is the ratio $ \epsilon /T$, so
- on very general grounds - we expect this ratio to by the correct
expression for the entropy per particle, irrespective whether the
particle is compound or elementary and whether it is relativistic
or non-relativistic.

Note that the proposed relation $\sigma = \epsilon / T$ for the
entropy per particle fits very well with the definition of
entropy: Assume that we have a (compound) particle with total
energy $\epsilon$ residing in a thermal bath with temperature $T$.
Now add some energy (in the form of "heat" $\delta Q$) to the
system and transfer this energy in a reversible
process\footnote{Note that any microscopic process is reversible.}
to the compound particle. Leave the temperature of the thermal
bath unchanged. For a molecule this could be achieved by inducing
a vibrational or rotational mode. For a black hole this could be
the absorption of a photon. The total energy of the compound
particle will change. According to the first law of thermodynamics
$\delta \epsilon = \delta Q$. However, the entropy of the particle
must have changed as well according to the definition of the
entropy: $\delta S = \delta Q / T = \delta \epsilon / T$. There is
nothing - in principle - that could prevent us to make the
internal energy change as large as we want. In the ultimate limit
$\sigma_m = \epsilon/T$. According to the equivalence principle
this relation should hold in general, irrespective of the size or
nature of any particle.

However, if we express equation (\ref{eq:App:sigma:m}) in terms of
the total energy and temperature, we find:

\begin{equation}
\sigma_m =\frac{\epsilon}{T} + (1-u)
\end{equation}

There is an "unwanted" additional factor $1-u$. Our expectation
was wrong, unless $u=1$. The miraculous thing is, that $u=1$ is
{\em exactly} what is expected, when one takes into account that
the {\em free} energy $F = E - ST$ must be minimized for any
(closed) system. We will see later in section
\ref{app:free:energy} that - in the non-relativistic case - the
free energy of an ideal gas is minimized to zero, when the
chemical potential per temperature $u$ is unity.

Furthermore, in \cite{petri/hol} it has been shown, that the so
called holographic solution of general relativity is compatible
with the Hawking entropy-area law {\em only if} $F = 0$ throughout
the whole interior of the solution. $F=0$ is a general
requirement, which not only holds in the non-relativistic
(matter-dominated) low energy phase, but also at
ultra-relativistic energies.

This can easily be seen: If $\sigma =  \epsilon / T$, then $F=0$
follows automatically. One just has to multiply $\sigma$ and
$\epsilon$ with the particle number to obtain $S = E/T$, which
implies $F= E - S T = 0$.

\subsubsection{\label{app:energy:pp}The energy per particle}

The energy per particle $\epsilon = e / n$ follows from equations
(\ref{eq:App:e}, \ref{eq:App:n}):

\begin{equation}
\epsilon = m + \eta_\epsilon \, T
\end{equation}

with

\begin{equation}
\eta_\epsilon = \frac{ \int{n_P \, z \, I' \, dz} }{\int{n_P \, I' \, dz} }
\end{equation}

For the non-relativistic case $\eta_\epsilon$ can be approximated by:

\begin{equation}
\eta_\epsilon \rightarrow \frac{\int_0^\infty{ e^{-z}
z^\frac{3}{2} dz} }{\int_0^\infty{ e^{-z} z^\frac{1}{2} dz}} =
\frac{\Gamma(\frac{5}{2})}{\Gamma(\frac{3}{2}) } = \frac{3}{2}
\end{equation}

so that

\begin{equation}
\epsilon = m + \frac{3}{2} \, T
\end{equation}

This is exactly as expected: Any translational degree of freedom
contributes a thermal energy of $T/2$ to the total energy of the
particle.

In the ultra-relativistic case we find:

\begin{equation}
\eta_\epsilon \rightarrow \frac{ Z_{3}(u) }{ Z_{2}(u) } > 0
\end{equation}

so that

\begin{equation}
\epsilon = m + \eta_\epsilon \, T \rightarrow \frac{
Z_{3}(u) }{ Z_{2}(u)} \, T
\end{equation}

The energy per particle in the ultra-relativistic case is
proportional to the temperature. The factor or proportionality
only depends on the dimensionless variable $u = \mu / T$. If $u =
const$, i.e. if the chemical potential is a linear function of the
temperature, the mean energy per particle is strictly proportional
to the temperature.

\subsubsection{The free energy-density}

The free energy-density $f = F /V$ follows from the definition of
the free energy $F = E - S T$. We find:

\begin{equation}
f = e - s T = g_F T^4 \int_0^\infty{ n_P (u I' - I) dz}
\end{equation}

In the non-relativistic case we can approximate this relation as
follows:

\begin{equation}
f \rightarrow g \sqrt{\frac{\pi}{2}} T^{\frac{5}{2}}
m^{\frac{3}{2}} e^{-(\frac{m}{T}-u)} (u-1)
\end{equation}

For a system at constant volume the free energy is minimized. In
the non-relativistic case the free energy is exactly zero for
$u=1$.\footnote{If one interprets the chemical potential per
temperature as an independent parameter, $f(u)$ as a function of
the single variable $u$ obtains its minimum value for $u=0$, as
can be seen easily by setting the derivative of $f \propto e^u
(1-u)$ with respect to $u$ to zero.}

In the ultra-relativistic case we find

\begin{equation}
f \rightarrow g_F T^4 \left( u Z_{2}(u) - \frac{Z_{3}(u)}{3}
\right)
\end{equation}

so that the free energy is minimized to zero whenever

\begin{equation}
Z_{3}(u) = 3 u Z_{2}(u)
\end{equation}

The above equation is an implicit equation for $u$, which can be
solved numerically. For fermions one finds $u \simeq \pm 1.34416$.
For an ideal gas consisting exclusively out of bosons there is no
solution for the above equation.

\subsubsection{\label{app:free:energy}The free energy per particle}

The free energy per particle $\phi = F / N$ is given by

\begin{equation}
\phi = \frac{f}{n} = T (u - \eta_\phi)
\end{equation}

with

\begin{equation}
\eta_\phi = \frac{ \int{n_P I dz} }{ \int{n_P I' dz} }
\end{equation}

In the non-relativistic case $\eta_\phi$ is equal to unity (see
section \ref{app:ideal:gas:law}), so that

\begin{equation}
\phi \rightarrow T (u - 1)
\end{equation}

As in the previous section, the free energy per particle is zero,
whenever $u=1$. This result is independent of the temperature.

In the ultra-relativistic case $\eta_\phi$ depends on the value of
the chemical potential per temperature:

\begin{equation} \label{eq:App:eta:phi}
\eta_\phi \rightarrow \frac{ Z_{3}(u) }{3 Z_{2}(u) }
\end{equation}

so that

\begin{equation}
\phi \rightarrow T (u - \frac{ Z_{3}(u) }{3 Z_{2}(u) })
\end{equation}

For $u=0$ one can express $\eta_\phi$ by the Riemann
$\zeta$-function (see section \ref{app:ideal:gas:law}). The
numerical values are: $\phi \rightarrow \eta_\phi \, T \simeq
-1.0505 \, T$ for fermions and $\phi \rightarrow \eta_\phi \, T
\simeq -0.90039 \, T$ for bosons.

\subsubsection{\label{app:ideal:gas:law}The ideal gas law}

Combining equations (\ref{eq:App:P}, \ref{eq:App:n}) we get

\begin{equation}
P = R \, n \, T
\end{equation}

with

\begin{equation}
R = \eta_\phi = \frac{\int_0^\infty{n_P \, I \, dz}}{\int_0^\infty{n_P \, I'
\, dz}}
\end{equation}

In the non-relativistic case it is easy to see from the
approximations given in section \ref{app:nonrel} that

$$R = \frac{\frac{2}{3} \Gamma(\frac{5}{2})}{\Gamma(\frac{3}{2})} = 1$$

We get the ideal gas law $P V = N T$ (in units $k=c=1 \rightarrow
R=1$).

In the ultra-relativistic case we have from equation (\ref{eq:App:eta:phi}):

$$R = \frac{Z_3(u)}{3 Z_2(u)}$$

which evaluates to $R_B = \pi^4 /(90 \zeta(3)) = 0.90039$ for
bosons and $R_F = 7/6 \, R_B \simeq 1.0505$ for fermions in the
case $u=0$.

\subsubsection{The ratio of particles to anti-particles}

Particles are labeled by a positive value of $u$ and
anti-particles with a negative value. For number-ratio we find
from equation (\ref{eq:App:n}):

$$\eta_N = \frac{ \int_0^\infty{n_P(u) I' dz} }{\int_0^\infty{n_P(-u) I' dz}}$$

In the ultra-relativistic case this reduces to

$$\eta_N = \frac{Z_2(u)}{Z_2(-u)}$$

At ultra-relativistic energies the bosons have $u=0$, therefore it
only makes sense to talk of a non-trivial ratio of particle to
anti-particle numbers for the fermions.

For non-relativistic energies we find, independent of particle
species:

$$\eta_N = e^{2u}$$

For $u=1$ we have $\eta_N \simeq 7.39$. Curiously this is quite
close to the ratio of protons to neutrons in our
universe.\footnote{Quite obviously the neutron is not the
anti-particle of the proton. Both particles are compound and have
different masses. Yet, assuming that the up and down quarks both
have $u \approx 1$ (as suggested from minimizing the free energy),
the proton, composed out of ($u u \overline{d}$) will have $u_p
\approx 1$, whereas the neutron ($u \overline{d} \overline{d}$)
has $u_n \approx -1$. Still, the equilibrium ratio of protons to
neutrons is governed by the term $\Delta m / T$, which is
formidable for the low temperatures encountered today.}

\subsection{Summary}

The thermodynamics of an ideal gas depends on the following
dimensionless quantities:

\begin{equation} \label{eq:Apps:x}
x = \frac{m}{T}
\end{equation}

\begin{equation} \label{eq:Apps:u}
u = \frac{\mu}{T}
\end{equation}

The following dimensionless quantity appears in the
Boltzmann-factor for the particle occupation number:

\begin{equation} \label{eq:Apps:alpha}
\alpha = \frac{\epsilon - \mu}{T} = z - \frac{\mu}{T} +
\frac{m}{T} = z - u + x
\end{equation}

The occupation number $n_P$ is given by:

\begin{equation}
n_P = \frac{1}{e^{\frac{\epsilon - \mu}{T}} \pm 1}=
\frac{1}{e^{\alpha} \pm 1} = \frac{1}{e^{z-u+x} \pm 1}
\end{equation}

where the plus-sign refers to fermions ($n_F$) and the minus sign
to bosons ($n_B$)

$z$ is a dimensionless integration variable: $z = (\epsilon -
m)/T$ with $\epsilon^2 = p^2 + m^2$.

Furthermore we need the following functions:

\begin{equation} \label{eq:Apps:I'}
I'(z, x) =  (z + x) \sqrt{z (z + 2 x)}
\end{equation}

and

\begin{equation} \label{eq:Apps:I}
I(z, x) =  \frac{\left(z (z + 2x)\right)^{\frac{3}{2}}}{3}
\end{equation}

The grand canonical potential density $j = J/V$ is given by:

\begin{equation} \label{eq:Apps:j}
j = -g T^4 \int_0^\infty{n_P \, I \, dz}
\end{equation}

where the constant $g$ depends linearly on the number of degrees
of freedom:

$$g = \frac{f}{2 \pi^2 \hbar^3}$$

The entropy-density $s = S/V$ is:

\begin{equation} \label{eq:Apps:s}
s = g T^3 \int_0^\infty{n_P \, \left(I + \alpha I' \right) \, dz}
\end{equation}

The number-density $n = N/V$

\begin{equation} \label{eq:Apps:n}
n = g T^3 \int_0^\infty{n_P \, I' \, dz}
\end{equation}

The energy-density $e = E/V$ is given by:

\begin{equation} \label{eq:Apps:e}
e = g T^4 \int_0^\infty{n_P \left( x + z\right) \, I' \, dz}
\end{equation}

and the (partial) pressure $P$:

\begin{equation} \label{eq:App:P=-j}
P = -j = g T^4 \int_0^\infty{n_P \, I \, dz}
\end{equation}

All other thermodynamic relations can be obtained by combining the
four equations above. Approximations for $I$, $I'$ etc. for the
non-relativistic and the ultra-relativistic case can be found in
sections \ref{app:rel} and \ref{app:nonrel}.

For bosons $x \geq u$ is a general requirement. For fermions no
such restriction applies.

\end{document}